\documentclass[english,pra,reprint,longbibliography,superscriptaddress]{revtex4-2}
\usepackage[utf8]{inputenc}
\usepackage{xcolor}
\usepackage{babel}
\usepackage{verbatim}
\usepackage{mathrsfs}
\usepackage{amsmath}
\usepackage{amssymb}
\usepackage{graphicx}
\usepackage{esint}
\usepackage[pdfusetitle,
 bookmarks=true,bookmarksnumbered=false,bookmarksopen=false,
 breaklinks=false,pdfborder={0 0 0},pdfborderstyle={},backref=false,colorlinks=true]
 {hyperref}
\hypersetup{
 hypertexnames=false}

\makeatletter

\providecommand{\tabularnewline}{\\}

\usepackage{graphicx}
\usepackage{dcolumn}
\usepackage{bm}
\usepackage{braket}
\usepackage{bbold}
\usepackage{tikz}
\usetikzlibrary{graphs, graphs.standard}
\usepackage{amsthm}
\usepackage{orcidlink}

\makeatother

\begin{document}
\title{Optimization for the propagation of a multiparticle quantum walk in
a one-dimensional lattice}
\author{Daer Feng\orcidlink{0000-0001-8881-3310}}
\affiliation{School of Physics, Sun Yat-Sen University, Guangzhou, Guangdong 510275,
China}
\affiliation{Institute of Physics, Chinese Academy of Sciences, Beijing 100190,
China}
\affiliation{School of Physical Sciences, University of Chinese Academy of Sciences,
Beijing 100049, China}
\author{Shengshi Pang\orcidlink{0000-0002-6351-539X}}
\email{pangshsh@mail.sysu.edu.cn}

\affiliation{School of Physics, Sun Yat-Sen University, Guangzhou, Guangdong 510275,
China}
\affiliation{Hefei National Laboratory, University of Science and Technology of
China, Hefei 230088, China}
\begin{abstract}
The quantum walk is a quantum counterpart of the classical random
walk that exhibits nonclassical behaviors and outperforms the classical
random walk in various aspects. It has been known that a single particle
can be propagated by a discrete-time quantum walk with a quadratic
time scaling in the variance of position distribution, beating the
linear time scaling in a classical random walk. In this paper, we
consider the discrete-time quantum walk for multiple particles in
a one-dimensional lattice, and investigate the optimization of the
joint coin state to enhance the spatial propagation of the particles
in the lattice. We study the asymptotic evolution of position distribution
for multiple particles in the long-time limit, and analytically optimize
the joint coin state to derive the maximum variance of the position
distribution between the particles after the evolution of the quantum
walk. An interesting result is that an optimized coin state always
possesses specific exchange symmetry which can be characterized by
a graph consisting of two disconnected complete subgraphs and the
exchange symmetry can significantly influence the position correlations
between the particles, showing the critical role of coin symmetry
in the propagation of multiple particles by the quantum walk. We further
study the entanglement of the optimized coin states to show the relation
of the coin correlations to the particle position distribution.
\end{abstract}
\maketitle

\section{Introduction}

The quantum walk is an interesting quantum protocol with distinctive
properties, such as quadratic speedup in the propagation of particles
\citep{konno2002quantum}, and exponentially fast hitting \citep{childs2003exponential,tang2018experimental2},
compared to its classical counterpart. These nonclassical behaviors
have motivated intense research interest in exploring the intrinsic
nature of the quantum walk and its potential applications \citep{venegas-andraca2012quantum,kadian2021quantum}.
So far, the quantum walk has been found to be a powerful tool in many
quantum computing and quantum information tasks, e.g., spatial search
\citep{portugal2013quantum,wong2016laplacian,shenvi2003quantum},
graph isomorphism testing \citep{rudinger2012noninteracting,douglas2008aclassical,gamble2010twoparticle,berry2011twoparticle2},
universal quantum computation \citep{childs2013universal,singh2021universal,childs2009universal},
and quantum teleportation \citep{yan2019generalized,li2019anew}.
Quantum walks can also simulate physical phenomena in condensed matter
physics such as Anderson localization and Bloch oscillations. Anderson
localization \citep{anderson1958absence}, the absence of diffusion
of a quantum mechanical state, can be produced in a quantum walk when
the walk is subjected to disorder \citep{crespi2013anderson2,mendes2019localizationdelocalization2}
or when the walk is inhomogeneous in the transition of a walker from
one site to the others dependent on both the coin state and the position
state \citep{shikano2010localization,konno2010localization}. The
relation between the localization and the degeneracy of the eigenstates
of a quantum walk is studied in Refs. \citep{inui2005localization,inui2004localization}.
Lin \emph{et al.} simulated the non-Hermitian topological Anderson
insulator experimentally by disordered photonic quantum walks \citep{lin2022observation}.

Quantum walks have been realized experimentally in a vast variety
of physical systems \citep{manouchehri2014physical} such as superconducting
qubits \citep{gong2021quantum,yan2019strongly}, optical lattices
\citep{preiss2015strongly}, photons \citep{schreiber2010photons,broome2010discrete,schreiber2011decoherence,aspuru-guzik2012photonic,peruzzo2010quantum,schreiber2012a2d,tang2018experimental,cardano2015quantum,defienne2016twophoton,qiang2016efficient,jeong2013experimental},
Bose-Einstein condensates \citep{xie2020topological,dadras2018quantum},
and trapped ions \citep{huertaalderete2020quantum,schmitz2009quantum,zahringer2010realization,xue2009quantum}.
Experimental realizations of quantum walks via photons \citep{pathak2007quantum,sansoni2012twoparticle}
and particularly the Hong-Ou-Mandel effect \citep{karczewski2019genuine,simon2020quantumclustered2}
exhibit different dynamics depending on the photon indistinguishability
and statistics. Vertex search in graphs and graph isomorphism testing
have been implemented on silicon photonic quantum walk processors
\citep{qiang2021implementing}.

Two types of formulations have been developed for quantum walks so
far: the discrete-time quantum walk and the continuous-time quantum
walk. The continuous-time quantum walk was proposed by Farhi and Gutmann
as a powerful computational model that provides exponential speedup
in penetrating a decision tree to outperform its classical counterpart
\citep{farhi1998quantum}. The discrete-time quantum walk was introduced
by Aharonov \emph{et al.} \citep{Aharonov1993}, and it differs from
the continuous-time quantum walk in that the walkers evolve with discrete
time steps, and extra degrees of freedom, coins, are introduced to
control the shift directions of the walkers at each step. The discrete-time
quantum walk is also found useful in different quantum algorithms
\citep{kempe2003quantum}. And it inspires many variations of quantum
walk models: the coin-flip operator can be position dependent \citep{ahmad2020onedimensional},
leading to the quantum-walk-based search algorithms \citep{portugal2013quantum};
the walker can possess more than one coin and the coins can be entangled
\citep{liu2009onedimensional,venegas-andraca2018quantum,venegas-andraca2005quantum};
etc. Extension of quantum walks to higher-dimensional lattices \citep{annabestani2010asymptotic,roldan2013ndimensional2},
entangled particles \citep{gamble2010twoparticle,berry2011twoparticle2},
and interacting particles \citep{gamble2010twoparticle,berry2011twoparticle2,li2013discretetime,alonso-lobo2018twoparticle,costa2019multiparticle2,cai2021multiparticle}
have also been explored.

The introduction of coin degrees of freedom has a substantial influence
on quantum walks, as the coins enlarge the dimension of the joint
Hilbert space of the particles and can be entangled or interacting
with each other, which gives more possibility and flexibility in various
quantum tasks. For example, the presence of coins may enhance the
distinguishing power of quantum walks on nonisomorphic graphs \citep{berry2011twoparticle2},
the Grover coin can enable the quantum-walk-based search algorithm
to reach the time complexity $O(\sqrt{N})$ for a database of size
$N$ \citep{shenvi2003quantum} in analogy to the well-known Grover
search algorithm \citep{grover1996afast,grover1997quantum}, the coin
symmetry can lead to different conditions for infinite hitting time
\citep{Krovi2006a,Krovi2006,Prabhu2022}, the correlations between
two coins can develop spatial correlations between the walkers \citep{orthey2019nonlocality},
and the conditional shift can generate entanglement between the coin
and position degrees of freedom \citep{abal2006quantum,goyal2010spatial,vallejo2019entropy}.
Moreover, Tregenna \emph{et al.} have analytically investigated the
general two-dimensional unitary coin operation for a single-particle
quantum walk and showed the effect of the coin operation on the asymmetry
of the walker position distribution \citep{tregenna2003controlling}.
The effect of decoherence in the coins of a quantum walk has also
been studied, showing that decoherence could slow down the spatial
spreading of the particles in a quantum walk and lead to approximately
linear growth of the position distribution, which is typically a classical
behavior \citep{brun2003quantum}.

A fascinating property of discrete-time quantum walks is that the
variance of the position distribution of a single-particle quantum
walk scales quadratically with time, compared to the linear time scaling
of position variance in the classical random walk \citep{kempe2003quantum},
implying a particle propagates much faster by quantum walk than by
the classical counterpart, which is essentially rooted in the interference
effect of different evolution paths in quantum walks \citep{Aharonov1993,knight2004}.
And the coin state can tune the interference effect of the quantum
walk and influence the position distribution of the particles \citep{tregenna2003controlling}.
In particular, Omar \emph{et al.} \citep{omar2006quantum} elucidate
that for a two-particle discrete-time quantum walk, the exchange symmetry
and entanglement of the joint coin states can significantly change
the position correlations of the particles and, as a consequence,
alter the average distance between the particles and the speed that
the particles are propagated by the quantum walk.

As the fast propagation of particles is a prominent advantage of quantum
walks over classical random walks and the coin states can change the
position distribution of particles as reviewed above, we consider
the following question in this paper to explore the limit of advantage
that quantum walk can reach in propagating particles: For an arbitrary
number of particles, what kind of joint coin state can maximize the
relative distance between the particles in a quantum walk? We quantify
the relative distance of multiple particles by the variance of the
position distribution between the particles, and use the asymptotic
analysis approach to study the evolution of a multiparticle discrete-time
quantum walk in a one-dimensional lattice. The position variance between
the particles is obtained asymptotically in the long-time limit, and
the joint coin state is further analytically optimized to derive the
maximum of position variance. An interesting result is that the optimized
coin states always possess specific partial exchange symmetry between
the particles and the symmetry can be characterized in a graphical
approach. The entanglement and the two-particle correlations of the
optimized coin states are also investigated in detail and illustrated
by numerical computation.

The paper is structured as follows. In Sec. \ref{sec:mathematical-framework},
we provide preliminaries for the discrete-time quantum walk and the
extension to the multiparticle case. Section \ref{sec:scaling-of-multi-particle}
studies the evolution of multiparticle quantum walks for an arbitrary
number of particles and obtains the asymptotic position variance between
the particles to characterize the propagation property of multiparticle
quantum walks. In Sec. \ref{sec:Entangled=000020coin}, we apply the
result of position variance to the case that the coins can be initially
entangled and analytically optimize the coin state to derive the bounds
of the position variance between the particles. Section \ref{sec:Symmetry-and-entanglement}
reveals the exchange symmetry of the optimized coin state and its
crucial role in the position variance of the particles, and explores
the entanglement and position correlations between the walkers. The
paper is finally concluded in Sec. \ref{sec:Conclusions}.

\global\long\def\evolve{\hat{U}}%
\global\long\def\shift{\hat{S}}%
\global\long\def\coinOperator{\hat{C}}%
\global\long\def\position{x}%
\global\long\def\spin{s}%
\global\long\def\K{K}%
\global\long\def\dspin{d}%
\global\long\def\movetoright{\hat{Q}}%
\global\long\def\evolven{\hat{U}_{n}}%
\global\long\def\stateAtTime#1{|\psi_{#1}\rangle}%
\global\long\def\distanceOperator{\hat{D}}%
\global\long\def\IA{I_{A2}}%
\global\long\def\IAM{I_{A1}}%
\global\long\def\IB{I_{B}}%
\global\long\def\IBM{I_{B_{1}}}%
\global\long\def\IAMP{I_{a}}%
\global\long\def\matrixTsquare{M}%
\global\long\def\matrixTsquareElement{m}%
\global\long\def\IAoscill{I_{Ao}}%
\global\long\def\IBoscill{I_{Bo}}%
\global\long\def\IAC{I_{AC}}%
\global\long\def\f{f}%
\global\long\def\Dmat{Z}%

\global\long\def\ketForCoin#1{\mathinner{|\!#1\rangle}}%
\global\long\def\braForCoin#1{\mathinner{\langle#1\!|}}%

\section{Preliminaries\protect\label{sec:mathematical-framework}}

In this section, we introduce preliminaries of multiparticle discrete-time
quantum walks and notations that will be used in this paper.

For discrete-time quantum walks in an infinite one-dimensional lattice,
particles are located on discrete sites and possess coins to determine
the shift directions of the particles at each time step. The Hilbert
space of single-particle quantum walks in a one-dimensional lattice
can be decomposed as $\mathscr{H}=\mathscr{H}_{\mathrm{position}}\otimes\mathscr{H}_{\mathrm{coin}}$,
in which a single-particle state at position $\position$ with the
coin upwards or downwards can be written as $|\position\rangle\otimes\ketForCoin{\uparrow}$
or $|\position\rangle\otimes\ketForCoin{\downarrow}$, respectively.
The unitary evolution of the particle is given by
\begin{equation}
\evolve=\shift\cdot(\hat{I}\otimes\coinOperator),
\end{equation}
where $\coinOperator$ is a coin operator that flips the coin and
$\shift$ is the shift operator that changes the position of the particle
dependent on its coin state. $\shift$ usually takes the following
form,
\begin{equation}
\shift=\movetoright\otimes\ketForCoin{\uparrow}\braForCoin{\uparrow}+\movetoright^{\dagger}\otimes\ketForCoin{\downarrow}\braForCoin{\downarrow},\label{eq:shift}
\end{equation}
where $\movetoright$ and $\movetoright^{\dagger}$ are the shift
operators in the one-dimensional lattice. If $|x\rangle$ denotes
the eigenstate of the position operator $\hat{x}$ with eigenvalue
$x$, then $\hat{Q}$ and $\hat{Q}^{\dagger}$ can be chosen as
\begin{equation}
\movetoright=\sum_{\position}|\position+1\rangle\langle\position|,\;\movetoright^{\dagger}=\sum_{\position}|\position-1\rangle\langle\position|.
\end{equation}

To generalize the quantum walk in a one-dimensional lattice to noninteracting
multiparticle cases, the evolution operator becomes the tensor product
of single-particle evolution operators; i.e., $\evolven=\evolve^{\otimes n}$
for $n$ particles. For an initial state as $\stateAtTime 0$, the
final state after $t$ steps of evolution is $\stateAtTime t=\text{\ensuremath{\hat{U}_{n}^{t}}}\stateAtTime 0.$
A general $n$-particle initial state for a discrete-time quantum
walk can be written as
\begin{equation}
|\psi_{0}\rangle=\sum a_{\position_{1}\spin_{1},...,\position_{n}\spin_{n}}|\position_{1}\spin_{1}\rangle\otimes\cdots\otimes|x_{n}\spin_{n}\rangle.\label{eq:initstate}
\end{equation}
Here $|x_{i}\rangle$ and $|s_{i}\rangle$ denote the eigenstates
of the position operator $\hat{x}$ and the Pauli operator $\hat{\sigma}_{z}$
of the $i$th particle. After $t$ steps of evolution, the final state
of the particles can be obtained by calculating the linear combination
of the evolved basis states, which are the tensor products of the
final states of single-particle walks,
\begin{equation}
|\psi_{t}\rangle=\sum a_{\position_{1}\spin_{1},...,\position_{n}\spin_{n}}\evolve^{t}|\position_{1}\spin_{1}\rangle\otimes\cdots\otimes\evolve^{t}|x_{n}\spin_{n}\rangle.
\end{equation}
The quantum walk with two particles was studied in Ref. \citep{omar2006quantum},
and the exchange symmetry of the two particles was shown to have significant
influence on the evolution. For more particles, Chandrashekar and
Busch \citep{chandrashekar2012quantum} considered the multiparticle
quantum walk initialized in an uncorrelated state. When the particles
are uncorrelated initially, the final state can be written as $\evolve^{t}|\phi_{1}\rangle\otimes\evolve^{t}|\phi_{2}\rangle\otimes\cdots\otimes\evolve^{t}|\phi_{n}\rangle$,
where $|\phi_{i}\rangle$ is the initial single-particle state of
the $i$th particle. But in this paper, we consider a more general
situation in which the initial state is in the form of Eq. \eqref{eq:initstate},
which generally describes distinguishable particles but can also describe
indistinguishable particles when it is invariant under arbitrary particle
exchange.

The coin operator can be an arbitrary SU(2) transformation in general;
i.e., regardless of a global phase,
\begin{equation}
\coinOperator=\left[\begin{array}{cc}
\sqrt{\rho} & \sqrt{1-\rho}e^{i\theta}\\
\sqrt{1-\rho}e^{i\varphi} & -\sqrt{\rho}e^{i(\varphi+\theta)}
\end{array}\right],
\end{equation}
where $0\le\rho\le1$ and $0\le\theta,\varphi\le\pi$. In the current
work, we choose the Hadamard operator to be the coin operator. 

\section{Position variance of the multiparticle quantum walk \protect\label{sec:scaling-of-multi-particle}}

In this section, we define the position variance between particles
and calculate the position variance of a general $n$-particle state
for a sufficiently large number of walk steps.

In a classical multiparticle random walk, the variance of the position
distribution between the particles grows linearly with the number
of time steps, as is proven in Appendix \ref{sec:Distance-of-multi-walker}.
In this paper, we first show that for a general multiparticle quantum
walk, the position variance can grow quadratically with the number
of time steps, which also holds true when the particles are initially
entangled and the different walk paths may interfere \citep{knight2004}.

The variance of the position distribution between particles is defined
as
\begin{equation}
\distanceOperator=\sum_{i}\left[\hat{\position}_{i}-\frac{1}{n}\sum_{j}\hat{\position}_{j}\right]^{2},\label{eq:Doriginal}
\end{equation}
where $\hat{x}_{i}$ represents the position of the $i$th particle,
and $\distanceOperator$ can be decomposed into two parts,
\begin{equation}
\distanceOperator=\frac{n-1}{n}\sum_{i}\hat{\position}_{i}^{2}-\frac{1}{n}\sum_{j\ne j^{\prime}}\hat{\position}_{j}\hat{x}_{j^{\prime}},\label{eq:Doperator}
\end{equation}
consisting of single-particle operators and two-particle operators.
Note that this definition considers only the position variance \emph{between}
the particles, excluding the position variance of each single particle
which is studied in Refs. \citep{ambainis2001onedimensional,konno2002quantum,kempe2003quantum}
and also scales quadratically with time. When $n=2$, the variance
$\hat{D}$ can be reduced to the squared distance between the two
particles in Ref. \citep{omar2006quantum}.

\subsection{Diagonalization of the evolution operator}

Diagonalizing the evolution operator of the particles will help simplify
the calculation of the evolution. Following the idea of discrete-time
Fourier transform for quantum walks \citep{brun2003quantum,ambainis2001onedimensional},
one can transform the state of the particles from the position space
into the momentum space, and the transformation between the two spaces
is given by 
\begin{equation}
|\K\rangle=\frac{1}{\sqrt{2\pi}}\sum_{x}e^{i\K\position}|\position\rangle,
\end{equation}
where $|\K\rangle$ and $|x\rangle$ are the basis states of the momentum
space and the position space, respectively. It can be verified that
$|\K\rangle$ preserves the orthonormality $\langle\K^{\prime}|\K\rangle=\delta(\K^{\prime}-\K)$
and is the eigenstate of $\movetoright$ with eigenvalue $e^{-i\K}$
and the eigenstate of $\movetoright^{\dagger}$ with eigenvalue $e^{i\K}$;
therefore, applying the conditional shift $\shift$ given in Eq. \eqref{eq:shift}
to $|\K\rangle$ merely induces extra phases in the coin space, and
the evolution $\hat{U}$ of the particles in the joint Hilbert space
of the position and coin can be factorized as
\begin{equation}
\hat{U}=\int_{-\pi}^{\pi}dK|\K\rangle\langle K|\otimes\evolve_{\K},\label{eq:pos_diag_U}
\end{equation}
where $\evolve_{\K}=(e^{-i\K}\ketForCoin{\uparrow}\braForCoin{\uparrow}+e^{i\K}\ketForCoin{\downarrow}\braForCoin{\downarrow})\coinOperator$
is the conditional evolution in the coin space with a given momentum
$K$. If we choose the coin-flip operator to be the Hadamard operator
\begin{equation}
\hat{H}=\frac{1}{\sqrt{2}}\left(\ketForCoin{\uparrow}\braForCoin{\uparrow}+\ketForCoin{\uparrow}\braForCoin{\downarrow}+\ketForCoin{\downarrow}\braForCoin{\uparrow}-\ketForCoin{\downarrow}\braForCoin{\downarrow}\right),
\end{equation}
then $\evolve_{\K}$ can be written as
\begin{equation}
\evolve_{\K}=\frac{e^{-i\K}}{\sqrt{2}}\left(\ketForCoin{\uparrow}\braForCoin{\uparrow}+\ketForCoin{\uparrow}\braForCoin{\downarrow}\right)+\frac{e^{i\K}}{\sqrt{2}}\left(\ketForCoin{\downarrow}\braForCoin{\uparrow}-\ketForCoin{\downarrow}\braForCoin{\downarrow}\right).
\end{equation}
Further diagonalization of the evolution $\evolve_{\K}$ gives the
eigenstates as
\begin{equation}
\begin{aligned} & |\dspin_{1}(\K)\rangle=-\frac{e^{-i\K}}{\sqrt{2N(\K)}}|\uparrow\rangle+\frac{1}{\sqrt{2N(\pi-\K)}}|\downarrow\rangle,\\
 & |\dspin_{2}(\K)\rangle=\frac{e^{-i\K}}{\sqrt{2N(\pi-\K)}}|\uparrow\rangle+\frac{1}{\sqrt{2N(\K)}}|\downarrow\rangle,
\end{aligned}
\end{equation}
with $\lambda_{j}\,(j=1,2)$ as the corresponding eigenvalues,
\begin{equation}
\lambda_{j}(\K)=(-1)^{j}\frac{1}{2}\sqrt{3+\cos(2\K)}-\frac{i\sin(\K)}{\sqrt{2}},
\end{equation}
where $N(\K)=(1+\cos^{2}\K)+\cos\K\sqrt{1+\cos^{2}\K}.$

Finally, we arrive at the diagonalized representation of the momentum-dependent
unitary evolution operator $\hat{U}_{K}$ in Eq. \eqref{eq:pos_diag_U},
\begin{equation}
\evolve_{\K}=\lambda_{1}|d_{1}(\K)\rangle\langle\dspin_{1}(\K)|+\lambda_{2}|d_{2}(\K)\rangle\langle\dspin_{2}(\K)|.\label{eq:diagonal=000020evolution=000020operator}
\end{equation}

\subsection{Average position variance of the multiparticle walk\vspace{-2pt}}

As calculating the exact position variance between particles is generally
complex, we will focus on the case for a sufficiently large number
of time steps in this paper. We outline the method to compute the
asymptotic variance below, and leave the details of derivation to
the Appendix \ref{sec:Calculation-of-variance}.

For a general $n$-particle quantum walk with initial state given
in Eq. \eqref{eq:initstate}, the average position variance of the
particles after $t$ steps of evolution is $\langle\psi_{0}|(\evolven^{\dagger})^{t}\distanceOperator\evolven^{t}|\psi_{0}\rangle$,
and it can be broken down into the summation of the terms in the following
form:
\begin{equation}
\begin{aligned} & \langle\position_{1}^{\prime}\spin_{1}^{\prime}|\cdots\langle x_{n}^{\prime}\spin_{n}^{\prime}|(\evolven^{\dagger})^{t}\distanceOperator\evolven^{t}|\position_{1}\spin_{1}\rangle\cdots|x_{n}\spin_{n}\rangle\\
= & \frac{n-1}{n}\sum_{i}\langle\position_{i}^{\prime}\spin_{i}^{\prime}|(\evolve^{\dagger})^{t}\hat{\position}_{i}^{2}\evolve^{t}|\position_{i}\spin_{i}\rangle\prod_{\gamma=1,\gamma\ne i}^{n}\delta_{\position_{\gamma}^{\prime}\spin_{\gamma}^{\prime},\position_{\gamma}\spin_{\gamma}}\\
 & -\frac{1}{n}\sum_{j\ne k}\langle\position_{j}^{\prime}\spin_{j}^{\prime}|(\evolve^{\dagger})^{t}\hat{\position}_{j}\evolve^{t}|\position_{j}\spin_{j}\rangle\langle\position_{k}^{\prime}\spin_{k}^{\prime}|(\evolve^{\dagger})^{t}\hat{\position}_{k}\evolve^{t}|\position_{k}\spin_{k}\rangle\\
 & \times\prod_{\gamma=1,\gamma\ne j,k}^{n}\delta_{\position_{\gamma}^{\prime}\spin_{\gamma}^{\prime},\position_{\gamma}\spin_{\gamma}}.
\end{aligned}
\label{eq:crossing=000020terms=000020expand}
\end{equation}
$\delta_{\position_{\gamma}^{\prime}\spin_{\gamma}^{\prime},\position_{\gamma}\spin_{\gamma}}$
is $1$ when $x_{\gamma}^{\prime}=x_{\gamma}$, $s_{\gamma}^{\prime}=s_{\gamma}$
and $0$ otherwise, implying the summands in the second line can contribute
to the position variance only when the states of the particles other
than the $i$th particle in the bra $\langle\position_{1}^{\prime}\spin_{1}^{\prime}|\cdots\langle x_{n}^{\prime}\spin_{n}^{\prime}|$
and the ket $|\position_{1}\spin_{1}\rangle\cdots|x_{n}\spin_{n}\rangle$
match. And a similar constraint applies to the summands in the third
and fourth lines. For the matrix element $\langle\position_{i}^{\prime}\spin_{i}^{\prime}|(\evolve^{\dagger})^{t}\hat{\position}_{i}^{2}\evolve^{t}|\position_{i}\spin_{i}\rangle$,
by the completeness of the basis $\{|\K\rangle\otimes|\dspin\rangle\}$
and noting that Eqs. \eqref{eq:pos_diag_U} and \eqref{eq:diagonal=000020evolution=000020operator}
give the spectral decomposition of the evolution operator, we can
obtain
\begin{equation}
\begin{aligned}\langle\position_{i}^{\prime}\spin_{i}^{\prime}| & (\evolve^{\dagger})^{t}\hat{\position}_{i}^{2}\evolve^{t}|\position_{i}\spin_{i}\rangle\\
= & \int_{-\pi}^{\pi}\int_{-\pi}^{\pi}{\rm d}\K^{\prime}{\rm d}\K\sum_{\dspin,\dspin^{\prime}}\langle\position_{i}^{\prime}\spin_{i}^{\prime}|\K^{\prime}\dspin^{\prime}\rangle\\
 & \times\langle\K^{\prime}\dspin^{\prime}|(\evolve^{\dagger})^{t}\hat{\position}_{i}^{2}\evolve^{t}|\K\dspin\rangle\langle\K\dspin|\position_{i}\spin_{i}\rangle\\
= & \int_{-\pi}^{\pi}\int_{-\pi}^{\pi}{\rm d}\K^{\prime}{\rm d}\K\sum_{\dspin,\dspin^{\prime}}\\
 & \times\langle\position_{i}^{\prime}|\K^{\prime}\rangle\langle\spin_{i}^{\prime}|\dspin^{\prime}\rangle\langle\K|\position_{i}\rangle\langle\dspin|\spin_{i}\rangle\langle\dspin^{\prime}|(\evolve_{K^{\prime}}^{\dagger})^{t}|\dspin^{\prime}\rangle\\
 & \times\langle\K^{\prime}|\hat{\position}_{i}^{2}|\K\rangle\langle\dspin^{\prime}|\dspin\rangle\langle\dspin|\evolve_{K}^{t}|\dspin\rangle.
\end{aligned}
\label{eq:matrix=000020element=000020ut=000020x^2=000020ut}
\end{equation}
Representing $\hat{x}^{2}$ in the momentum space produces the periodic
extension of the derivative of the Dirac delta function with period
$2\pi$,
\begin{equation}
\begin{aligned}\langle\K^{\prime}|\hat{\position}_{i}^{2}|\K\rangle= & \sum_{\position=-\infty}^{\infty}\langle\K^{\prime}|\hat{\position}_{i}^{2}|\position\rangle\langle\position|\K\rangle\\
= & -\sum_{l=-\infty}^{\infty}\delta^{(2)}(\K^{\prime}-\K+2\pi l),
\end{aligned}
\end{equation}
where $l\in\mathbb{Z}$. So, the matrix elements of $(\evolve^{\dagger})^{t}\hat{\position}_{i}^{2}\evolve^{t}$
can be simplified to
\begin{equation}
\begin{aligned}\langle\position_{i}^{\prime}\spin_{i}^{\prime}| & (\evolve^{\dagger})^{t}\hat{\position}_{i}^{2}\evolve^{t}|\position_{i}\spin_{i}\rangle\\
= & -\frac{1}{2\pi}\int_{-\pi}^{\pi}\int_{-\pi}^{\pi}{\rm d}\K^{\prime}{\rm d}\K\sum_{\dspin,\dspin^{\prime}}e^{i\K^{\prime}x_{i}^{\prime}}e^{-iKx_{i}}\langle\spin_{i}^{\prime}|\dspin^{\prime}\rangle\langle\dspin|\spin_{i}\rangle\\
 & \times\langle\dspin^{\prime}|(\evolve_{K^{\prime}}^{\dagger})^{t}|\dspin^{\prime}\rangle\delta^{(2)}(\K^{\prime}-\K)\langle\dspin^{\prime}|\dspin\rangle\langle\dspin|\evolve_{K}^{t}|\dspin\rangle\\
= & -\frac{1}{2\pi}[t^{2}\IA(\position_{i}^{\prime}-\position_{i},\spin_{i}^{\prime},\spin_{i})+t\IAM(\position_{i}^{\prime},\position_{i},\spin_{i}^{\prime},\spin_{i})\\
 & +\IAC(\position_{i}^{\prime},\position_{i},\spin_{i}^{\prime},\spin_{i})+\IAoscill(\position_{i}^{\prime},\position_{i},\spin_{i}^{\prime},\spin_{i};t)].
\end{aligned}
\label{eq:single=000020body}
\end{equation}
Similarly,
\begin{equation}
\begin{aligned} & \langle\position_{j}^{\prime}\spin_{j}^{\prime}|(\evolve^{\dagger})^{t}\hat{\position}_{j}\evolve^{t}|\position_{j}\spin_{j}\rangle\\
= & -\frac{i}{2\pi}[t\IB(\position_{j}^{\prime}-\position_{j},\spin_{j}^{\prime},\spin_{j})+\IBM(\position_{j}^{\prime},\position_{j},\spin_{j}^{\prime},\spin_{j})\\
 & +\IBoscill(\position_{j}^{\prime},\position_{j},\spin_{j}^{\prime},\spin_{j};t)].
\end{aligned}
\label{eq:two=000020body}
\end{equation}

In the above equations, $\IA$, $\IAM$, $\IAC$, $\IB$, and $\IBM$
are coefficients for different powers of $t$ and can be calculated
using the residue theorem. The integrals in the term $t^{2}$ of the
position variance are
\begin{equation}
\begin{aligned}\IA(\position,\spin^{\prime},\spin)= & \begin{cases}
(1-\sqrt{2})f(0), & \position=0,\spin^{\prime}=\spin\\
\f(x), & \position\ne0,\spin^{\prime}=\spin\\
0, & \spin^{\prime}\ne\spin
\end{cases}\end{aligned}
\end{equation}
\begin{equation}
\begin{aligned}\IB(\position,\spin^{\prime},\spin)= & \begin{cases}
(-1)^{\spin}i\IA(\position,\spin^{\prime},\spin), & \spin^{\prime}=s\\
\frac{i}{2}[f(x)+f(x-(-1)^{\spin}\times2)], & \spin^{\prime}\ne\spin,
\end{cases}\end{aligned}
\end{equation}
where $\f(x)=\sqrt{2}\pi\left(\sqrt{2}-1\right)^{|\position|}\cos\left(\pi x/2\right)$.
When $\spin$ is $\uparrow$, it denotes the number 1, and $\downarrow$
denotes $0$; i.e., when $\spin=\downarrow$, $(-1)^{\spin}=1$. The
detailed expressions of integrals $\IAM$, $I_{B_{1}}$, $\IAC$,
$\IAoscill$ and $\IBoscill$ are given in Appendix \ref{sec:Calculation-of-variance}.

$\IAoscill$ and $\IBoscill$ are integrals involving rapidly time-varying
phases when $t$ is large, which can be calculated by the stationary
phase approximation \citep{bleistein2010}. The matrix elements \eqref{eq:single=000020body}
and \eqref{eq:two=000020body} are the functions of the states $|\position_{i}\spin_{i}\rangle$
and $|\position_{i}^{\prime}\spin_{i}^{\prime}\rangle$ as well as
the number of time steps, $t$, with the highest order of $t$ being
$t^{2}$. After a long time evolution, the $t^{2}$ term becomes dominant,
so the other lower-order terms can be dropped. By plugging Eqs. \eqref{eq:single=000020body}
and \eqref{eq:two=000020body} into Eq. \eqref{eq:crossing=000020terms=000020expand}
and summing up all the terms in the form of Eq. \eqref{eq:crossing=000020terms=000020expand},
it yields the average position variance for a given initial state
in Eq. \eqref{eq:initstate}.

\section{Bounds of position variance\protect\label{sec:Entangled=000020coin}}

Now we consider the case in which  the particles are initially uncorrelated
in the spatial lattice but entangled in the coin space. We are concerned
with how the position variance between the particles is dependent
on the initial coin state, and we obtain the upper and lower bounds
of the position variance when the number of time steps is large.

\begin{figure}
\begin{centering}
\includegraphics[scale=0.62]{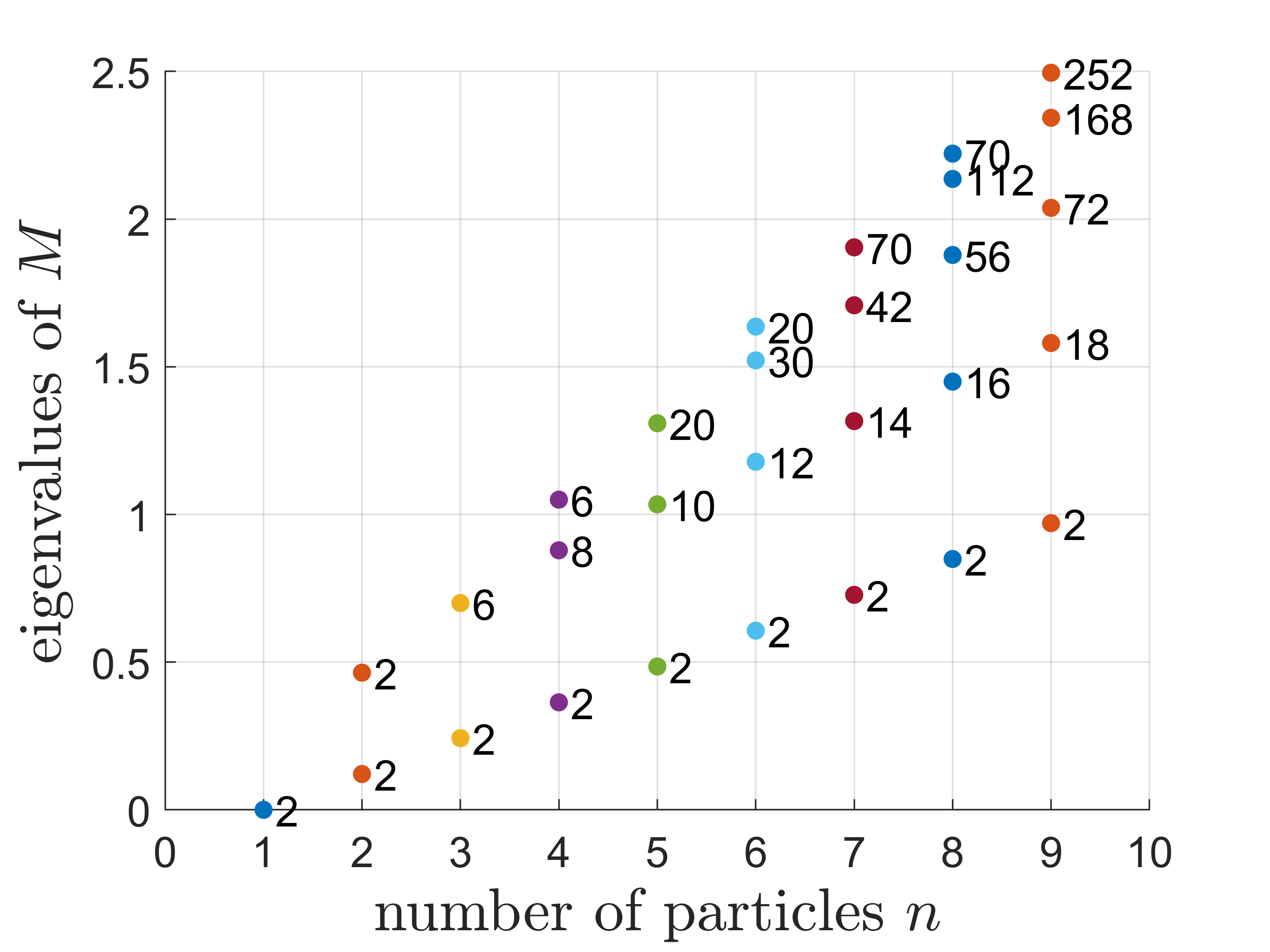}
\par\end{centering}
\caption{\protect\label{fig:Eigenvalues-and-their}Distribution of the eigenvalues
of matrix $M$ in Eq. \eqref{eq:M}. The eigenvalues are the extremized
coefficients of the leading time-quadratic term in the position variance,
for different numbers of particles. The numbers next to the points
denote the degrees of degeneracy of the eigenvalues.}
\end{figure}

To simplify the representation of the coin state, we denote the basis
states of the coin space, $\ketForCoin{\uparrow}$ and $\ketForCoin{\downarrow}$
as $|1\rangle$ and $|0\rangle$ and the coin state of the $i$th
particle as $|s_{i}\rangle$. The combination of the coin states of
all particles can be treated as a binary number and can be simply
denoted by its numerical value, i.e., $|\spin_{1}\spin_{2}\cdots\spin_{n}\rangle=|2^{n-1}\times\spin_{1}+2^{n-2}\times\spin_{2}+\cdots+2\spin_{n-1}+\spin_{n}\rangle=|\xi\rangle$,
where $\xi$ is the numerical value of the binary number $\spin_{1}\spin_{2}\cdots\spin_{n}$.
We assume the spatial state of the particles to be a product state,
and the initial state can be written as 
\begin{equation}
|\Psi\rangle=|x_{1}\rangle|x_{2}\rangle\cdots|x_{n}\rangle\otimes\sum_{\xi=0}^{2^{n}-1}a_{\xi}|\xi\rangle.\label{eq:inital=000020coin=000020state}
\end{equation}
For simplicity, we assume the particles to be located in the same
arbitrary position for the rest of this paper.

The position variance $\langle\Psi|(\evolve^{\dagger})^{t}\distanceOperator\evolve^{t}|\Psi\rangle$
after $t$ steps of evolution is generally complicated, but can be
expanded to a power series of $t$. For a large $t$, the $t^{2}$
term dominates the variance, so we will mainly consider the coefficient
of $t^{2}$ denoted as $c_{2}$ in the following. It can be verified
that $c_{2}$ has a quadratic form of $a_{0},a_{1},...,a_{2^{n}-1}$
and their complex conjugates, which can be rearranged in the following
matrix form: 
\begin{equation}
c_{2}=[a_{0}^{*},a_{1}^{*},...,a_{2^{n}-1}^{*}]\matrixTsquare[a_{0},a_{1},...,a_{2^{n}-1}]^{T},\label{eq:quadratic=000020form}
\end{equation}
where $M$ collects all the coefficients of $a_{i}^{*}a_{j}$ in $c_{2}$.
And it turns out that $c_{2}$ is independent of the initial positions
of particles in the state $|\Psi\rangle$ in Eq. \eqref{eq:inital=000020coin=000020state}.
The matrix $M$ is obtained in Appendix \ref{sec:Optimization-problem-of},
and the elements of $M$ are obtained as
\begin{equation}
\begin{aligned}\matrixTsquare_{ii}= & (n-1)\left(1-1/\sqrt{2}\right)\\
 & +\frac{1}{n}[n-(2W(i)-n)^{2}]\left(1-1/\sqrt{2}\right)^{2},\\
\matrixTsquare_{jk}= & \frac{2}{n}\left(1-1/\sqrt{2}\right)^{2}[\delta_{d(j,k),1}(n-1\\
 & -2\min\{W(j),W(k)\})-\delta_{d(j,k),2}],
\end{aligned}
\label{eq:M}
\end{equation}
where $i,j,k\in\{0,1,2,...,2^{n}-1\}$, $j\neq k$, $W(i)$ is the
Hamming weight of the binary representation of integer $i$, $d(j,k)$
is the Hamming distance between integers $j$ and $k$, i.e., the
number of bit positions where $k$ differs from $j$, and $\delta$
is the Kronecker delta function.

The maximum and minimum values of $c_{2}$ are exactly the maximum
and minimum eigenvalues of the matrix $M$ since the complex vector
$[a_{0},a_{1},...,a_{2^{n}-1}]$ is normalized, and the normalized
eigenvectors of the matrix $\matrixTsquare$ are the extremal points
for the coefficient $c_{2}$ {[}Eq. \eqref{eq:quadratic=000020form}{]}
and equivalently for the quadratic term of $t$ in the position variance
of $|\Psi\rangle$. By numerical computation, the relation of the
number of particles to the eigenvalues of $\matrixTsquare$ and their
degeneracies is illustrated in Fig. \ref{fig:Eigenvalues-and-their}.
And it is proven in Appendix \ref{sec:Solving-eigenvalues-and} that
the eigenvalues are
\begin{equation}
\eta_{k}=\left(1-\frac{1}{\sqrt{2}}\right)^{2}\left[\frac{8W(k)(n-W(k))}{n}+(n-1)\sqrt{2}\right],\label{eq:eigenvalue2}
\end{equation}
where $k=0,2,...,2^{n}-2$. It can be inferred that  $0\le W(k)\le n-1$,
$W(k)\in\mathbb{Z}$, and there are $C_{n-1}^{W(k)}$ different values
of $k$'s that correspond to the same value of $W(k)$. Note that
for different values of $k$'s, e.g., $k$ and $k^{\prime}$, $\eta_{k}$
and $\eta_{k^{\prime}}$ take the same value when $W(k)=n-W(k^{\prime})$,
which should be taken into account in counting the degeneracies of
the eigenvalues. The degeneracy of $\eta_{k}$ is
\begin{equation}
\begin{aligned} & \mathrm{count}(\eta_{k})\\
= & \begin{cases}
2, & k=0\\
2(C_{n-1}^{W(k)}+C_{n-1}^{n-W(k)}), & k\ne0,n\mathrm{\ is\ odd}\\
2(C_{n-1}^{W(k)}+C_{n-1}^{n-W(k)}), & k\ne0,W(k)\ne\frac{n}{2},n\mathrm{\ is\ even}\\
2C_{n-1}^{W(k)}, & W(k)=\frac{n}{2},n\mathrm{\ is\ even}.
\end{cases}
\end{aligned}
\end{equation}
The calculation of the degeneracy of eigenvalues is also given in
Appendix \ref{sec:Solving-eigenvalues-and}.

The eigenvectors for $\eta_{k}$ turn out to be the $k$th and $(k+1)$th
columns of matrix $P$, and $P$ is a $2^{n}\times2^{n}$ matrix consisting
of $2\times2$ submatrices 
\begin{equation}
P_{ik}=(-1)^{c(i,k)}\Dmat^{d(i,k)}=(-1)^{c(i,k)}\left[\begin{array}{cc}
0 & 1\\
1 & 2
\end{array}\right]^{d(i,k)},\label{eq:eigenvectors}
\end{equation}
where $i,k\in\{0,2,...,2^{n}-2\}$ are the row and column locations
of the upper left corner element of the submatrix $P_{ik}$ in the
matrix $P$. $c(i,k)$ denotes the number of bits on which the binary
form of $i$ and $k$ have a common $1$. The matrix $P$ diagonalizes
the matrix $M$ by $P^{-1}MP$. For instance, when $n=2$,\def\lenbrace{11}
\newcommand{\mata}{
\overbrace{\rule{\lenbrace mm}{0mm}}^{P_{00}}
}
\newcommand{\matb}{
\overbrace{\rule{\lenbrace mm}{0mm}}^{P_{02}}
}
\newcommand{\matc}{
\underbrace{\rule{\lenbrace mm}{0mm}}_{P_{20}}
}
\newcommand{\matd}{
\underbrace{\rule{\lenbrace mm}{0mm}}_{P_{22}}
}
\newcommand{\myarray}[3]{
\begin{array}{c}
	#1 \\[-2pt]
	#2 \\[-7pt]
	#3 \\
\end{array}
}
\begin{equation}
P=\myarray{\begin{array}{cc}
{\color{red}\mata} & {\color{blue}\matb}\end{array}}{\left[\begin{array}{cc|cc}
{\color{red}1\begin{array}{c}
\quad\end{array}} & {\color{red}0}\begin{array}{c}
\hphantom{}\end{array} & {\color{blue}0} & {\color{blue}1}\\
{\color{red}0\begin{array}{c}
\quad\end{array}} & {\color{red}1}\begin{array}{c}
\hphantom{}\end{array} & {\color{blue}1} & {\color{blue}2}\\
\hline {\color{violet}0}\begin{array}{c}
\quad\end{array} & {\color{violet}1}\begin{array}{c}
\hphantom{}\end{array} & \mathbin{\color{teal}-}{\color{teal}1} & {\color{teal}0}\\
{\color{violet}1}\begin{array}{c}
\quad\end{array} & {\color{violet}2}\begin{array}{c}
\hphantom{}\end{array} & {\color{teal}0} & \mathbin{\color{teal}-}{\color{teal}1}
\end{array}\right]}{\begin{array}{cc}
{\color{violet}\matc} & {\color{teal}\matd}\end{array}}.
\end{equation}
 Then $P^{-1}MP=\mathrm{diag}(\eta_{0},\eta_{0},\eta_{2},\eta_{2})=\mathrm{diag}(\frac{3}{\sqrt{2}}-2,\frac{3}{\sqrt{2}}-2,4-\frac{5}{\sqrt{2}},4-\frac{5}{\sqrt{2}})$.

By assigning the elements of column vectors in $P$ after normalization
to the corresponding coefficients $a_{\xi}$, we obtain optimized
coin states $\sum_{\xi=0}^{2^{n}-1}a_{\xi}|\xi\rangle$.

The eigenvalues of the matrix $\matrixTsquare$ give the range of
$c_{2}$, $\eta_{\mathrm{min}}\le c_{2}\le\eta_{\mathrm{max}}$. According
to Eq. \eqref{eq:eigenvalue2}, the largest eigenvalue for a given
number of particles $n$ is
\begin{equation}
\eta_{\mathrm{max}}(n)=\begin{cases}
\left(1-\frac{1}{\sqrt{2}}\right)^{2}[(n-1)\sqrt{2}+2n], & n\ \mathrm{even},\\
\left(1-\frac{1}{\sqrt{2}}\right)^{2}[(n-1)\sqrt{2}+2n-\frac{2}{n}], & n\ \mathrm{odd},
\end{cases}
\end{equation}
while the smallest eigenvalue is
\begin{equation}
\eta_{\mathrm{min}}(n)=\left(1-\frac{1}{\sqrt{2}}\right)^{2}\sqrt{2}(n-1).
\end{equation}
The eigenvalues $\eta_{\mathrm{max}}(n)$ and $\eta_{\mathrm{min}}(n)$
are positive, so both of the upper bound and lower bound of the position
variance grow quadratically with the number of time steps $t$. The
eigenvectors for $\eta_{\mathrm{min}}(n)$ are the columns $k=0$
and $k=1$ of the matrix $P$. And when $n$ is odd, the eigenvectors
for $\eta_{\mathrm{max}}(n)$ are the columns $k$ and $k+1$ where
$k$ satisfies $W(k)=\lceil n/2\rceil$ or $W(k)=\lfloor n/2\rfloor$;
when $n$ is even, the eigenvectors for $\eta_{\mathrm{max}}(n)$
are the columns $k$ and $k+1$ where $k$ satisfies $W(k)=n/2$.

The above solution to the eigenvalue problem of the matrix $M$ in
the quadratic form of $c_{2}$ provides the optimized initial coin
states and the extremal values for the leading term of $t$ in the
position variance of the $n$-particle state. For simplicity, we refer
to these optimized coin states as eigenstates, and sometimes we refer
to an eigenstate as the initial particle state combining both the
optimized coin state and the positional state as is shown in Eq. \eqref{eq:inital=000020coin=000020state}.

\section{Symmetry and entanglement of optimized coin states \protect\label{sec:Symmetry-and-entanglement}}

Next, we proceed to study the relation between the extremal values
of the coefficient $c_{2}$ for the quadratic term of $t$ in the
particle position variance and the symmetry of the initial coin states
of the $n$ particles. As we will see later, each eigenstate is invariant
after arbitrary permutation among a subset of the particles. We define
this invariance of the eigenstates as \emph{partial exchange symmetry}.

\subsection{Partial exchange symmetry of optimized coin states}

\begin{figure}
\begin{centering}
\includegraphics{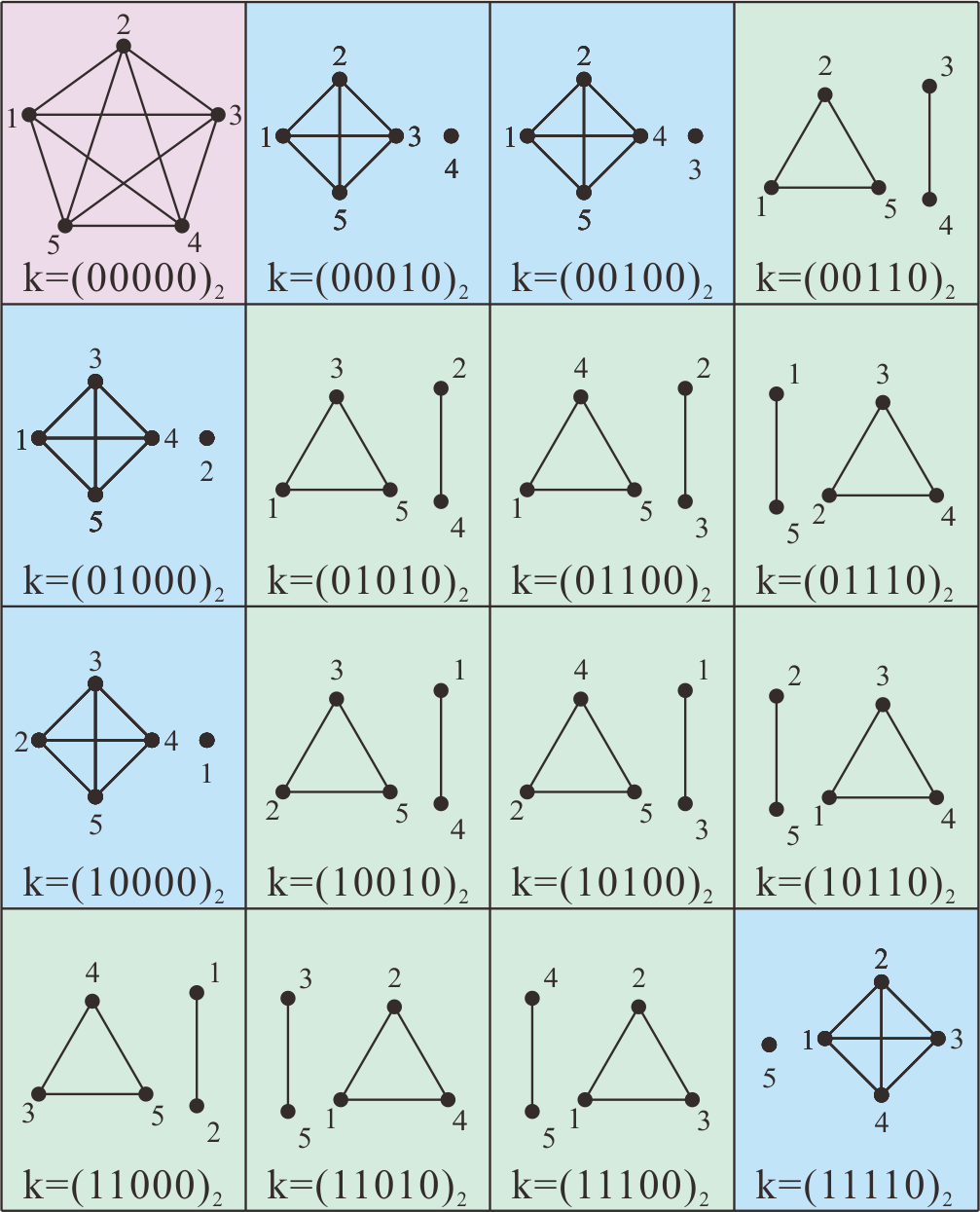}
\par\end{centering}
\caption{\protect\label{fig:completegraphs}Vertex partitions for eigenstates
of $n=5$ particles. The numbers near the vertices label different
particles, and the subset that the $i$th particle belongs to depends
on the $i$th bit of $k$. From left to right and then top to bottom,
each cell of the table represents complete graph(s) corresponding
to a different even number $k$ ranging from $0$ to $2^{n}-2$, respectively,
where $k$ is the column subscript of the corresponding eigenvector
in the matrix $P$. When $k$ is odd, the vertex partition is the
same as that of $k-1$. Different background colors of the cells are
assigned to the graphs with different structures, which also indicate
different eigenvalues.}
\end{figure}

In Eq. \eqref{eq:inital=000020coin=000020state}, the positional state
is an arbitrary product state to make the results general, but it
has been shown above that the initial positions of the particles do
not affect the leading term $c_{2}t^{2}$ in the variance of the particle
positions. For convenience, we assume all of the particles to be located
at the original point initially, so that the symmetry of the particles
is only determined by the joint coin state. If we visualize the partial
exchange symmetry of an eigenstate by assigning the particles to the
vertices of an undirected graph and connecting a pair of particles
with an edge if the state is unchanged after swapping those two particles,
it turns out that the graph may be composed of two disconnected but
complete subgraphs. This is illustrated in Fig. \ref{fig:completegraphs},
and proven in Appendix \ref{sec:Proof-of-partial}.

It is shown that the number of possible ways $p_{k}$ to exchange
two particles that preserve an eigenstate is related to the eigenvalue
$\eta_{k}$ by $\eta_{k}=\left(1-1/\sqrt{2}\right)^{2}\left[8(C_{n}^{2}-p_{k})/n+(n-1)\sqrt{2}\right]$,
so one can deduce that the larger the $p_{k}$ is, the smaller the
eigenvalue $\eta_{k}$ will be, and thus a greater degree of partial
exchange symmetry of the eigenstate results in a smaller relative
spreading distance of the quantum walk, which is essentially induced
by a greater extent of constructive interference between the walking
routes of the particles and more particles oriented in the same direction
and thus closer to each other.

It can be proven that (see Appendix \ref{sec:Proof-of-partial}) if
one denotes $n_{\uparrow}$ and $n_{\downarrow}$ as the  numbers
of ``ones'' and ``zeros'' in the binary form of $k$ ($k$ is
even) respectively, then $p_{k}=\frac{n_{\uparrow}(n_{\uparrow}-1)}{2}+\frac{n_{\downarrow}(n_{\downarrow}-1)}{2}$,
$n_{\uparrow}+n_{\downarrow}=n$, and the two disconnected subgraphs
have $n_{\uparrow}$ and $n_{\downarrow}$ vertices respectively.
When $k=0$, all the $n$ particles are assigned to the same subgraph
(actually the whole graph is connected in this case), which corresponds
to the largest $p_{k}$ and the smallest $\eta_{k}$. The particles
are closest to each other in this case and the eigenstate possesses
the greatest exchange symmetry. When the particles are distributed
as uniformly as possible between the two subgraphs, we obtain the
smallest $p_{k}$ and the particles turn out to be the most distant.
Note there is no antisymmetric eigenstate here when $n>2$ since the
coin space of each particle is only two dimensional, in contrast to
the case of two particles where both symmetric and antisymmetric coin
states can exist but lead to drastically different position distributions
\citep{omar2006quantum}.

It is worth noting that the framework of multiparticle quantum walks
starts with distinguishable particles in this paper, but the above
optimization results can actually also apply to indistinguishable
or partially indistinguishable particles. While the nature of indistinguishable
particles is intrinsically different from that of distinguishable
particles, the states of fully or partially indistinguishable particles
can be equivalently represented by fully or partially symmetric states
of distinguishable particles. So in mathematics, the quantum walk
of multiple indistinguishable particles can also be included in the
current framework with proper symmetric multiparticle states. The
above optimization of multiparticle position variance further implies
that the states of distinguishable particles possess full or partial
symmetry when the position variance is extremized, so the optimized
states of distinguishable particles coincide with the states of fully
or partially indistinguishable particles and the optimization results
can thus be applicable to indistinguishable particles as well.

\begin{figure}
\begin{centering}
\includegraphics[scale=0.7]{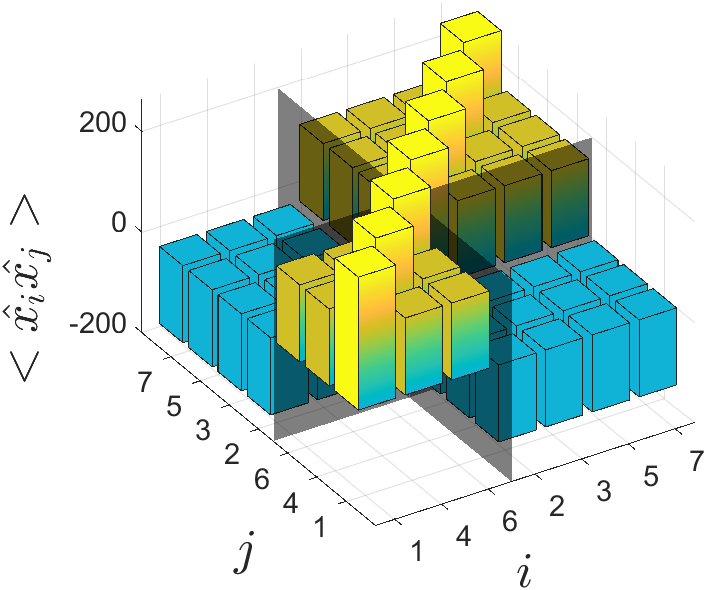}
\par\end{centering}
\caption{\protect\label{fig:The-average-of}Two-particle position correlations
of the evolved eigenstate with $k=(1001010)_{2}$, $n=7$, $t=30$.
$i$ and $j$ label the $i$th and $j$th particles, respectively.
The diagonal bars correspond to the average of $\hat{x}_{i}^{2}$,
while the off-diagonal bars represent two-particle position correlations.
There are three types of two-particle combinations: both particles
from the same subgraph (there are two disconnected subgraphs and thus
two different situations) or one particle from each subgraph. Therefore,
the three-dimensional bar graph is divided into four regions by two
vertical planes. The evolution does not break the exchange symmetry
of particles in the subgraphs; thus the correlations between two arbitrary
particles from the same subgraph remain the same.}
\end{figure}

It should be clarified that the bipartite structures of the graphs
in Fig. \ref{fig:completegraphs} do not mean that the two parts of
the graph are not correlated. For the eigenstate associated with the
$k$th eigenvector, if $k\ne0$, the entanglement between the two
disconnected parts of the graph can be quantified by the von Neumann
entropy 
\begin{equation}
S(\rho_{\uparrow})=-\sum_{i=1,2}\nu_{i}{\rm \log}\nu_{i},
\end{equation}
in which $\nu_{1}=[1+(3-2\sqrt{2})^{n-2}]^{-1}$ and $\nu_{2}=[1+(3+2\sqrt{2})^{n-2}]^{-1}$
for an even $k$, and $\nu_{1}=[1+(3-2\sqrt{2})^{n}]^{-1}$ and $\nu_{2}=[1+(3+2\sqrt{2})^{n}]^{-1}$
for an odd $k$ (Appendix \ref{sec:Entanglement-between-complete}).
$\rho_{\uparrow}$ denotes the reduced density matrix of the particles
associated with $n_{\uparrow}$. Interestingly, the results of $\nu_{1}$
and $\nu_{2}$ rely on the parity of $k$ only, independent of the
specific value of $k$ when $k$ is restricted to be odd or even $(k\ne0)$.
As $S(\rho_{\uparrow})$ is nonzero, the two disconnected parts of
the graphs are generally entangled for the variance of the particle
positions to reach extremal values.

\subsection{Dynamics of optimized coin states}

\begin{figure*}

\includegraphics[scale=0.98]{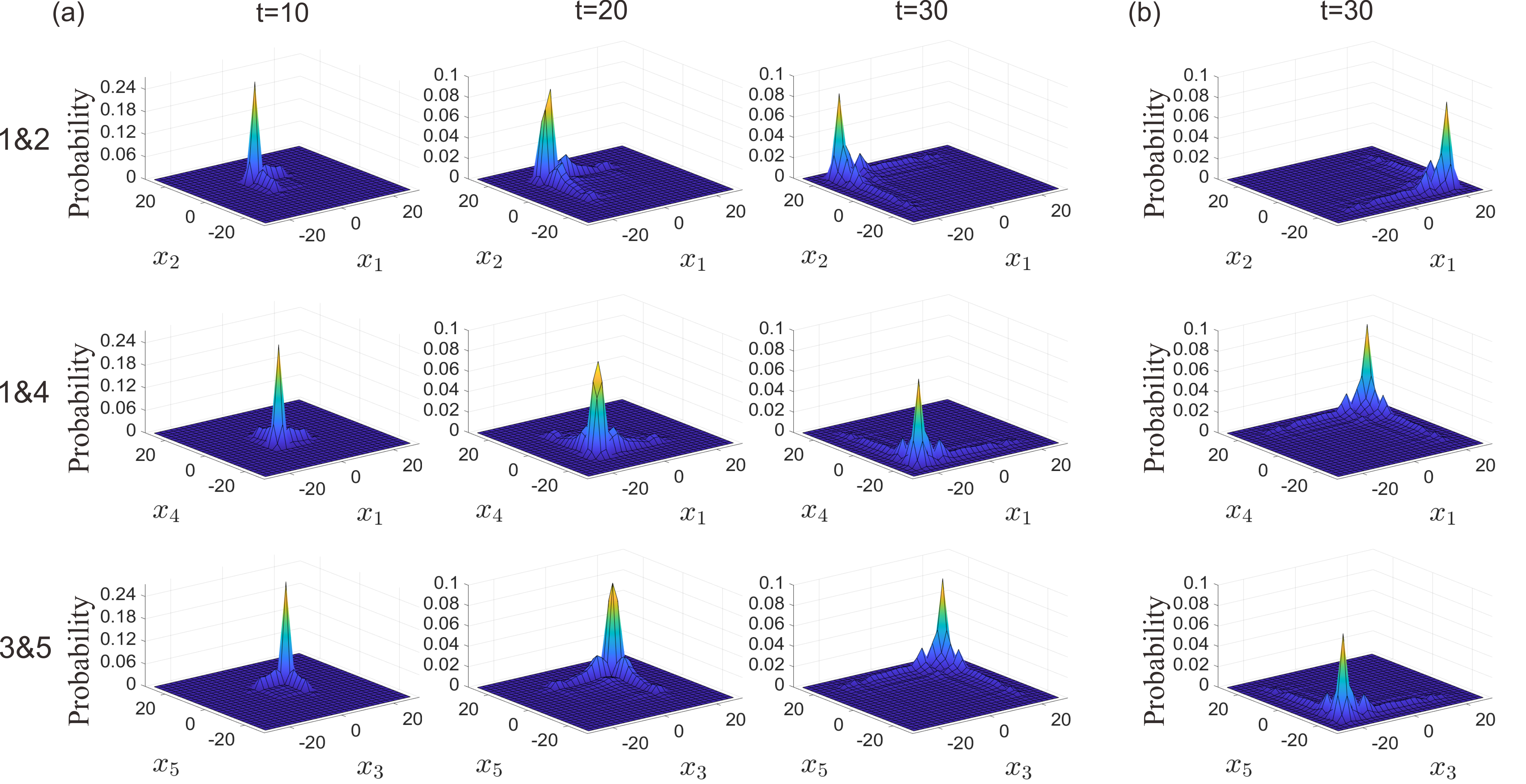}

\caption{\protect\label{fig:Two-particle-probability-distrib}(a) Two-particle
position distributions of particles 1 and 2, 1 and 4, and 3 and 5,
in the eigenstate with $k=(1001010)_{2}$, evolved for $t=10,20$,
and $30$ steps, respectively. For particles 1 and 2, as they belong
to different subgraphs, they are more likely to be found in different
positions, particularly particle 1 in negative positions and particle
2 in positive positions, since the coins are on average oriented in
the opposite directions as can be seen from the initial reduced density
matrix of the coins. For two particles in the same subgraph, i.e.,
1 and 4, and 3 and 5, on account of the exchange symmetry among particles
within the same subgraph, they are close to each other. (b) Two-particle
position distributions in the eigenstate associated with $k=(0110100)_{2}$
after $t=30$ steps of evolution.}

\end{figure*}

Before concluding this paper, we try to dig a bit deeper into the
dynamics of the particles in the two subgraphs. The initial positions
of all particles are assumed to be zero.

For the case of $n=7$, the average of $\hat{x}_{i}^{2}$ and two-particle
position correlations of the eigenstate given by the column $k=(1001010)_{2}$
of the matrix $P$ after an evolution of $t=30$ steps are shown in
Fig. \ref{fig:The-average-of}. They are essentially the components
of the position variance operator in Eq. \eqref{eq:Doperator}. It
can be seen that the particles in the same subgraph possess the same
average value of $\hat{x}_{i}^{2}$ and the same particle position
correlation due to the partial exchange symmetry and the evolution
does not break this symmetry. The position correlations of two particles
are positive if the particles are from the same subgraph and negative
if from different subgraphs, because the particles from the two subgraphs
are initially placed at the original point but oriented in different
directions by their coins. The two-particle position correlations
always remain the same if either particle is replaced by another particle
in the same subgraph, since picking different particles from the same
subgraph does not make any difference due to the partial exchange
symmetry.

\begin{figure}
\begin{centering}
\includegraphics[scale=0.65]{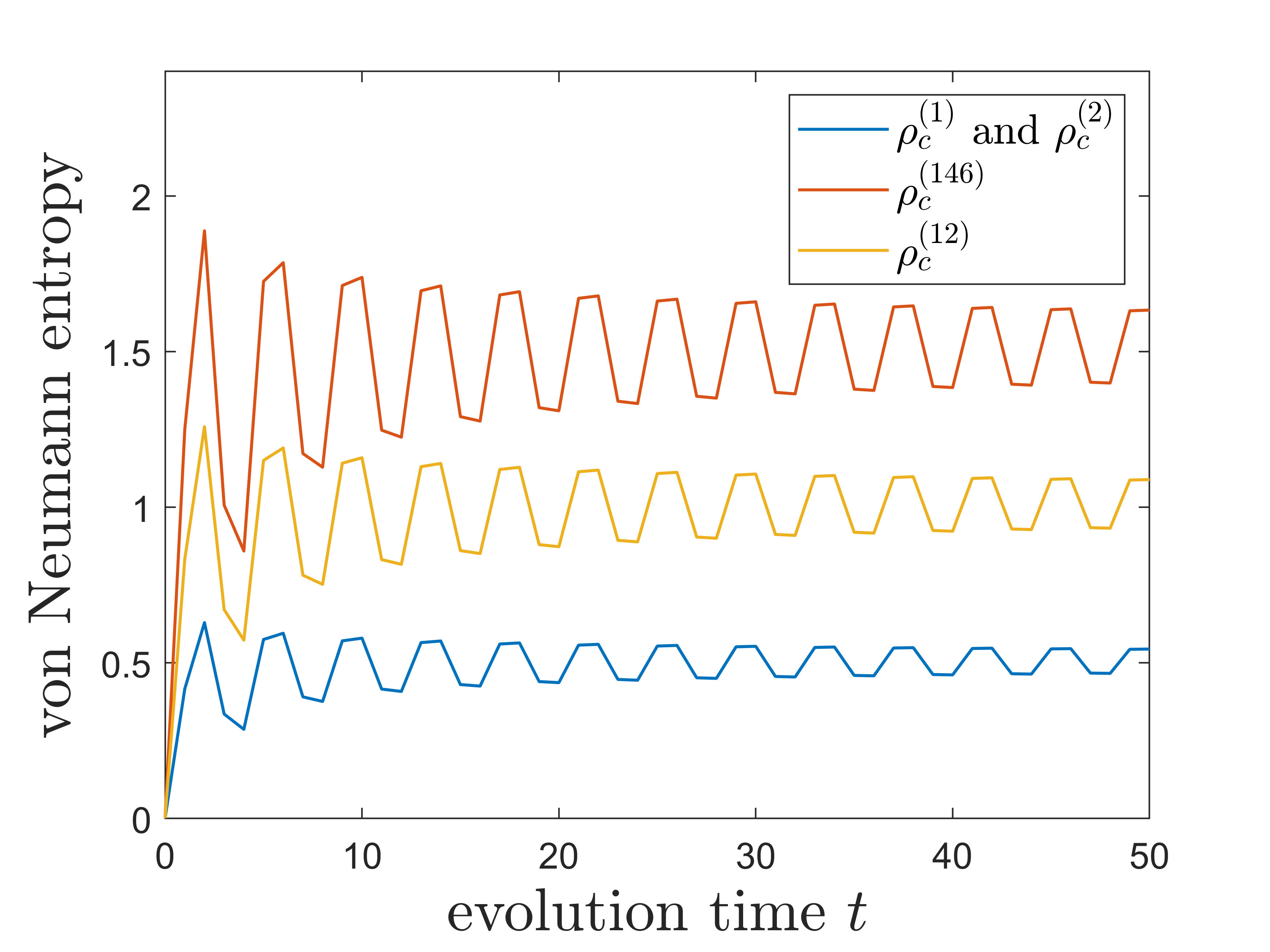}
\par\end{centering}
\caption{\protect\label{fig:Bipartite-entanglement}Bipartite entanglement
between the coins denoted in the legend and all the other degrees
of freedom of the seven particles over $t=50$ steps of evolution.
For instance, the red line plots the von Neumann entropy for the reduced
density matrix $\rho_{c}^{(146)}$ of the coins of particles $1$,
$4$, and $6$. The blue line and the yellow line likewise.}
\end{figure}

A more insightful observation of the position distribution of the
particles is presented in Fig. \ref{fig:Two-particle-probability-distrib}.
Particles in the same subgraph are more likely to be close to each
other. But particles from different subgraphs, such as particles 1
and 2, are more distant. As is shown in Fig. \ref{fig:Two-particle-probability-distrib}(a),
particle 1 is more likely to be found in negative positions while
particle 2 is more likely to be found in positive positions, because
we can see from the reduced density matrices of the initial coin state
that the coins are on average oriented in the opposite directions.
As a comparison, Fig. \ref{fig:Two-particle-probability-distrib}(b)
illustrates the position distributions of the particles in the evolved
eigenstate with $k=(0110100)_{2}$ after $t=30$ steps of evolution.
The Hamming weights of $(1001010)_{2}$ and $(0110100)_{2}$ are identical,
and $(0110100)_{2}$ can be perceived as relabeling particles $1,4$,
and $6$ as particles $2,3$ and $5$ and particles $2,3$, and $5$
as particles $1,4$, and $6$ in $(1001010)_{2}$. Referring to the
expression of $P_{ik}$ {[}Eq. \eqref{eq:eigenvectors}{]}, the two
eigenstates are essentially the same after the relabeling and therefore
have the same position variance, particularly the same second-order
term of $t$.

The bipartite entanglement between part of coins and all the other
degrees of freedom of the seven particles in the initial state $k=(1001010)_{2}$
is small, compared with that of the evolved state, as shown in Fig.
\ref{fig:Bipartite-entanglement}. We use the von Neumann entropy
as the entanglement measure. The entanglement increases rapidly at
the beginning and then starts to oscillate.

\section{Conclusions\protect\label{sec:Conclusions}}

In this paper, we studied the discrete-time multiparticle quantum
walk in a one-dimensional lattice, focusing on the position distribution
of the particles. We are interested in how fast the relative distance
between the particles can increase with time which is a manifestation
of the nonclassicality of the quantum walk over its classical counterpart.
We used the variance of the position distribution of the particles
to quantify the average relative distance between the particles after
the evolution of quantum walk, and obtained an asymptotic result of
the position variance for a sufficiently large number of time steps.

To explore the limit of spatial propagation of the particles in a
quantum walk, we analytically optimized the coin state and derived
the upper and lower bounds of the position variance for multiple particles
that are initially uncorrelated in the position space but allowed
to be entangled in the coin space, and both bounds turn out to scale
quadratically with the number of time steps, which is much faster
than the linear time scaling in the classical case.

Interesting symmetry in the optimized multiparticle coin states is
revealed for the the dominant term of position variance: the optimized
coin states that extremize the dominant term of position variance
possess partial exchange symmetry with respect to two disjoint subsets
of the particles; i.e., the optimized coin states are invariant under
arbitrary permutation of the particles in either subset. The two subsets
of particles in the optimized coin states are correlated, and we studied
the entanglement between the two subsets quantified by the von Neumann
entropy. We also investigated the position correlations and position
distributions of the optimized coin states to reveal more dynamical
properties of the quantum walk induced by the partial exchange symmetry.
\begin{acknowledgments}
The authors acknowledge the helpful discussions with Litong Deng from
Tsinghua University and Professor Xiaoyong Fu from Sun Yat-sen University.
This work is supported by the National Natural Science Foundation
of China (Grant No. 12075323), the Innovation Program for Quantum
Science and Technology (Grant No. 2021ZD0300702), and the College
Students Innovation and Entrepreneurship Training Program, Sun Yat-sen
University. 
\end{acknowledgments}


\appendix

\section*{Appendix: Details of the derivation}

\global\long\def\evolve{\hat{U}}%
\global\long\def\shift{\hat{S}}%
\global\long\def\coinOperator{\hat{C}}%
\global\long\def\position{x}%
\global\long\def\spin{s}%
\global\long\def\K{K}%
\global\long\def\dspin{d}%
\global\long\def\movetoright{\hat{Q}}%
\global\long\def\evolven{\hat{U}_{n}}%
\global\long\def\stateAtTime#1{|\psi_{#1}\rangle}%
\global\long\def\distanceOperator{\hat{D}}%
\global\long\def\IA{I_{A2}}%
\global\long\def\IAM{I_{A1}}%
\global\long\def\IB{I_{B}}%
\global\long\def\IBM{I_{B_{1}}}%
\global\long\def\IAMP{I_{a}}%
\global\long\def\matrixTsquare{M}%
\global\long\def\matrixTsquareElement{m}%
\global\long\def\IAoscill{I_{Ao}}%
\global\long\def\IBoscill{I_{Bo}}%
\global\long\def\IAC{I_{AC}}%
\global\long\def\f{f}%

\global\long\def\ketForCoin#1{\mathinner{|\!#1\rangle}}%
\global\long\def\braForCoin#1{\mathinner{\langle#1\!|}}%

\global\long\def\F{F}%
\global\long\def\B{B}%
\global\long\def\f{f}%

In this Appendix, we give more details about the derivation of results
in the main text. Section \ref{sec:Distance-of-multi-walker} calculates
the position variance of multiwalker classical random walks. Section
\ref{sec:Diagonalization-of-the} illustrates the preliminaries of
discrete-time multiparticle quantum walks and process of diagonalizing
the evolution operator. Section \ref{sec:Calculation-of-variance}
derives the position variance of the walkers as a function of the
evolution time $t$ and initial state. We consider a situation where
the particles can be entangled in the coin space, and express the
coefficient $c_{2}$ of $t^{2}$ into matrix multiplication in Sec.
\ref{sec:Optimization-problem-of}. Section \ref{sec:Solving-eigenvalues-and}
provides an approach to derive the eigenvalues and eigenvectors of
matrix $M$. We analyze and prove the partial exchange symmetry of
the particles of the eigenstates corresponding to the eigenvectors
in Sec. \ref{sec:Proof-of-partial}, and then in Sec. \ref{sec:Entanglement-between-complete}
we calculate the bipartite entanglement between the two subsets of
particles in which the particles within the same subset possess exchange
symmetry.

\renewcommand\thesubsection{\arabic{subsection}}
\setcounter{equation}{0}
\renewcommand\theequation{A\arabic{equation}}

\subsection{Position variance of multiwalker classical random walk\protect\label{sec:Distance-of-multi-walker}}

Considering a classical random walk with $n$ independent walkers,
we define the position variance of the particles as
\begin{equation}
\bar{D}=\overline{\sum_{i=1}^{n}\left[x_{i}-\frac{1}{n}\sum_{j=1}^{n}x_{j}\right]^{2}}=\frac{n-1}{n}\sum_{i}\overline{x_{i}^{2}}-\frac{1}{n}\sum_{j\ne k}\bar{x}_{j}\bar{x}_{k}.
\end{equation}
$x_{i}=x_{i0}+x_{i1}+\cdots+x_{it}$ is the final position of the
$i$th particle. $x_{it}\in\{-1,1\}$ is the displacement of the $i$th
particle at step $t$ ($t>0$), and the probability of $x_{it}$ to
be $-1$ or $1$ is equal. $x_{i0}$ is the initial position of the
$i$th particle. Therefore, $\bar{x}_{it}=0$ and $\overline{x_{it}^{2}}=1$:
\begin{equation}
\begin{aligned}\bar{D}= & \frac{n-1}{n}\sum_{i}\overline{(x_{i0}+x_{i1}+x_{i2}+\cdots+x_{it})^{2}}-\frac{1}{n}\sum_{j\ne k}\bar{x}_{j}\bar{x}_{k}\\
= & \frac{n-1}{n}\sum_{i}\overline{\left(\sum_{\gamma=0}^{t}\sum_{\gamma^{\prime}=0}^{t}x_{i\gamma}x_{i\gamma^{\prime}}\right)}-\frac{1}{n}\sum_{j\ne k}\bar{x}_{j}\bar{x}_{k}\\
= & \frac{n-1}{n}\sum_{i}\left[x_{i0}^{2}+2x_{i0}\sum_{\gamma=1}^{t}\bar{x}_{i\gamma}+\sum_{\gamma=1}^{t}\sum_{\gamma^{\prime}=1,\gamma^{\prime}\ne\gamma}^{t}\bar{x}_{i\gamma}\bar{x}_{i\gamma^{\prime}}\right.\\
 & \left.+\sum_{\gamma=1}^{t}\overline{x_{i\gamma}^{2}}\right]-\frac{1}{n}\sum_{j\ne k}x_{j0}x_{k0}\\
= & \frac{n-1}{n}\sum_{i}x_{i0}^{2}-\frac{1}{n}\sum_{j\ne k}x_{j0}x_{k0}+(n-1)t.
\end{aligned}
\end{equation}
$\bar{D}$ grows linearly as the function of $t$.

As for the position variance operator for quantum states, the simplification
of Eq. \eqref{eq:Doriginal} to Eq. \eqref{eq:Doperator} is carried
out as follows:
\begin{equation}
\begin{aligned}\hat{D}= & \sum_{i}\left[\hat{x}_{i}-\frac{1}{n}\sum_{j}\hat{x}_{j}\right]^{2}\\
= & \sum_{i}\left[\hat{x}_{i}^{2}-2\hat{x}_{i}\left(\frac{1}{n}\sum_{j}\hat{x}_{j}\right)+\left(\frac{1}{n}\sum_{j}\hat{x}_{j}\right)^{2}\right]\\
= & \sum_{i}\hat{x}_{i}^{2}-2\sum_{i}\hat{x}_{i}\left(\frac{1}{n}\sum_{j}\hat{x}_{j}\right)+\sum_{i}\left(\frac{1}{n}\sum_{j}\hat{x}_{j}\right)^{2}\\
= & \sum_{i}\hat{x}_{i}^{2}-2n\left(\frac{1}{n}\sum_{i}\hat{x}_{i}\right)\left(\frac{1}{n}\sum_{j}\hat{x}_{j}\right)+n\left(\frac{1}{n}\sum_{j}\hat{x}_{j}\right)^{2}\\
= & \sum_{i}\hat{x}_{i}^{2}-\frac{1}{n}\sum_{j,j^{\prime}}\hat{x}_{j}\hat{x}_{j^{\prime}}\\
= & \frac{n-1}{n}\sum_{i}\hat{x}_{i}^{2}-\frac{1}{n}\sum_{j\ne j^{\prime}}\hat{x}_{j}\hat{x}_{j^{\prime}}.
\end{aligned}
\end{equation}

\subsection{Diagonalization of the evolution operator\protect\label{sec:Diagonalization-of-the}}

The $n$-particle evolution operator is 
\begin{equation}
\evolven=\evolve^{\otimes n},
\end{equation}
and $\evolve=\shift\cdot(\hat{I}\otimes\coinOperator)$, where $\coinOperator$
is the coin operator that flips the coin and $\shift$ is the shift
operator that changes the position of the particle according to its
coin state. The coin of a particle is also referred to as “spin” wherever
no ambiguity occurs. $\shift$ takes the following form: $\shift=\sum_{\position}|\position+1\rangle\langle\position|\otimes\ketForCoin{\uparrow}\braForCoin{\uparrow}+\sum_{\position}|\position-1\rangle\langle\position|\otimes\ketForCoin{\downarrow}\braForCoin{\downarrow}$.

We perform a discrete-time Fourier transform
\begin{equation}
|\K\rangle=\frac{1}{\sqrt{2\pi}}\sum_{x}e^{i\K\position}|\position\rangle,\label{eq:DTFT-1}
\end{equation}
where $|\K\rangle$ preserves the orthonormality $\langle\K^{\prime}|\K\rangle=\delta(\K^{\prime}-\K)$,
to diagonalize the positional part of the evolution operator in the
momentum space that
\begin{equation}
\hat{U}|\K\rangle\otimes|\spin\rangle=|\K\rangle\otimes(e^{-i\K}\ketForCoin{\uparrow}\braForCoin{\uparrow}+e^{i\K}\ketForCoin{\downarrow}\braForCoin{\downarrow})\coinOperator|\spin\rangle.
\end{equation}
The calculation of the evolution can be converted into applying the
operator $\evolve_{\K}\equiv(e^{-i\K}\ketForCoin{\uparrow}\braForCoin{\uparrow}+e^{i\K}\ketForCoin{\downarrow}\braForCoin{\downarrow})\coinOperator$
in the coin space. As a specific case, we choose the Hadamard operator
$\hat{H}$ as the coin operator:
\begin{equation}
\hat{H}=\frac{1}{\sqrt{2}}\left(\ketForCoin{\uparrow}\braForCoin{\uparrow}+\ketForCoin{\uparrow}\braForCoin{\downarrow}+\ketForCoin{\downarrow}\braForCoin{\uparrow}\right)-\frac{1}{\sqrt{2}}\ketForCoin{\downarrow}\braForCoin{\downarrow}.
\end{equation}
Then $\evolve_{\K}$ can be written as
\begin{equation}
\evolve_{\K}=\frac{e^{-i\K}}{\sqrt{2}}\left(\ketForCoin{\uparrow}\braForCoin{\uparrow}+\ketForCoin{\uparrow}\braForCoin{\downarrow}\right)+\frac{e^{i\K}}{\sqrt{2}}\left(\ketForCoin{\downarrow}\braForCoin{\uparrow}-\ketForCoin{\downarrow}\braForCoin{\downarrow}\right).
\end{equation}
By diagonalizing it in the coin space further on, denoting $|\dspin_{i}(\K)\rangle$
as the eigenvectors of the new vector space, the diagonal representation
of operator $\evolve_{\K}$ is shown in Eq. \eqref{eq:diagonalrep},
where $|\dspin_{1}\rangle$ and $|\dspin_{2}\rangle$ are the two
eigenvectors.
\begin{equation}
\begin{aligned}\evolve_{\K}= & \frac{e^{-i\K}\left(-\sqrt{1+6e^{2i\K}+e^{4i\K}}-e^{2i\K}+1\right)}{2\sqrt{2}}|d_{1}\rangle\langle d_{1}|\\
 & +\frac{e^{-i\K}\left(\sqrt{1+6e^{2i\K}+e^{4i\K}}-e^{2i\K}+1\right)}{2\sqrt{2}}|d_{2}\rangle\langle d_{2}|\\
= & \lambda_{1}|d_{1}\rangle\langle d_{1}|+\lambda_{2}|d_{2}\rangle\langle d_{2}|.
\end{aligned}
\label{eq:diagonalrep}
\end{equation}
The actual values of $\lambda_{1}$ and $\lambda_{2}$ vary if we
choose different branches of the square-root function. Here we define
\begin{align}
\begin{aligned}\sqrt{1+6e^{2i\K}+e^{4i\K}}= & \sqrt{2}\cos(\K)\sqrt{\cos(2\K)+3}\\
 & +i\sqrt{2}\sin(K)\sqrt{\cos(2K)+3},
\end{aligned}
\end{align}
then
\begin{equation}
\begin{aligned}\lambda_{1}(\K)= & -\frac{1}{2}\sqrt{3+\cos(2\K)}-\frac{i\sin(\K)}{\sqrt{2}},\\
\lambda_{2}(\K)= & +\frac{1}{2}\sqrt{3+\cos(2\K)}-\frac{i\sin(\K)}{\sqrt{2}},
\end{aligned}
\end{equation}
\begin{equation}
|\dspin_{1}\rangle=-\frac{e^{-i\K}}{\sqrt{2N(\K)}}\ketForCoin{\uparrow}+\frac{1}{\sqrt{2N(\pi-\K)}}\ketForCoin{\downarrow},
\end{equation}
\begin{equation}
|\dspin_{2}\rangle=\frac{e^{-i\K}}{\sqrt{2N(\pi-\K)}}\ketForCoin{\uparrow}+\frac{1}{\sqrt{2N(\K)}}\ketForCoin{\downarrow},
\end{equation}
where $N(\K)=(1+\cos^{2}\K)+\cos\K\sqrt{1+\cos^{2}\K}=[\sqrt{1+\cos^{2}\K}+\cos\K]\sqrt{1+\cos^{2}\K}$.
\onecolumngrid Finally, we arrive at the diagonalized representation
of the evolution operator, 
\begin{equation}
\begin{aligned}\hat{U}= & \int_{-\pi}^{\pi}{\rm d}\K|\K\rangle\langle\K|\otimes(\lambda_{1}|d_{1}(\K)\rangle\langle\dspin_{1}(\K)|+\lambda_{2}|d_{2}(\K)\rangle\langle\dspin_{2}(\K)|).\end{aligned}
\label{eq:diagonal=000020evolution=000020operator-1}
\end{equation}

\subsection{Calculation of position variance\protect\label{sec:Calculation-of-variance}}

As is mentioned in the article, the initial state of an $n$-particle
quantum walk can be written as
\begin{equation}
|\psi_{0}\rangle=\sum_{\position_{1}\spin_{1}\cdots\position_{n}\spin_{n}}a_{\position_{1}\spin_{1}\cdots\position_{n}\spin_{n}}|\position_{1}\spin_{1}\rangle|x_{2}\spin_{2}\rangle\cdots|x_{n}\spin_{n}\rangle,
\end{equation}
where $x_{i}$ and $s_{i}$ are the position and spin of the $i$th
particle.

And the position variance operator is defined as
\begin{equation}
\distanceOperator=\sum_{i}\left[\hat{\position}_{i}-\frac{1}{n}\sum_{j}\hat{\position}_{j}\right]^{2}=\frac{n-1}{n}\sum_{i}\hat{\position}_{i}^{2}-\frac{1}{n}\sum_{j\ne j^{\prime}}\hat{\position}_{j}\hat{x}_{j^{\prime}},\label{eq:Doperator-1}
\end{equation}
consisting of single-particle operators and two-particle operators.
Similar to evolving operators in the Heisenberg picture, the evolved
position variance operator can be decomposed into the summation of
single-particle operators and two-particle operators:
\begin{equation}
\begin{aligned}(\evolve_{n}^{\dagger})^{t}\distanceOperator\evolve_{n}^{t}= & \frac{n-1}{n}\sum_{i}(\evolve_{i}^{\dagger})^{t}\hat{\position}_{i}^{2}\evolve_{i}^{t}-\frac{1}{n}\sum_{j\ne k}(\evolve_{j}^{\dagger})^{t}\hat{x}_{j}\evolve_{j}^{t}\otimes(\evolve_{k}^{\dagger})^{t}\hat{x}_{k}\evolve_{k}^{t}.\end{aligned}
\end{equation}
The subscripts $i,j,k$ represent which particle the operators apply
on. The position variance after $t$ steps of evolution is given by
\begin{equation}
\begin{aligned}\langle\psi_{0}|(\evolve_{n}^{\dagger})^{t}\distanceOperator\evolve_{n}^{t}|\psi_{0}\rangle= & \sum_{\position_{1}^{\prime}\spin_{1}^{\prime}\cdots\position_{n}^{\prime}\spin_{n}^{\prime}}\sum_{\position_{1}\spin_{1}\cdots\position_{n}\spin_{n}}a_{\position_{1}^{\prime}\spin_{1}^{\prime}\cdots\position_{n}^{\prime}\spin_{n}^{\prime}}^{*}a_{\position_{1}\spin_{1}\cdots\position_{n}\spin_{n}}[\frac{n-1}{n}\sum_{i}\langle\position_{i}^{\prime}\spin_{i}^{\prime}|(\evolve^{\dagger})^{t}\hat{\position}_{i}^{2}\evolve^{t}|\position_{i}\spin_{i}\rangle\prod_{\gamma=1,\gamma\ne i}^{n}\delta_{\position_{\gamma}^{\prime}\spin_{\gamma}^{\prime},\position_{\gamma}\spin_{\gamma}}\\
 & -\frac{1}{n}\sum_{j\ne k}\langle\position_{j}^{\prime}\spin_{j}^{\prime}|(\evolve^{\dagger})^{t}\hat{\position}_{j}\evolve^{t}|\position_{j}\spin_{j}\rangle\langle\position_{k}^{\prime}\spin_{k}^{\prime}|(\evolve^{\dagger})^{t}\hat{\position}_{k}\evolve^{t}|\position_{k}\spin_{k}\rangle\prod_{\gamma=1,\gamma\ne j,k}^{n}\delta_{\position_{\gamma}^{\prime}\spin_{\gamma}^{\prime},\position_{\gamma}\spin_{\gamma}}].
\end{aligned}
\end{equation}
$\delta_{\position_{\gamma}^{\prime}\spin_{\gamma}^{\prime},\position_{\gamma}\spin_{\gamma}}$
is $1$ if $\position_{\gamma}^{\prime}=x_{\gamma}$ and $\spin_{\gamma}^{\prime}=\spin_{\gamma}$,
and is $0$ otherwise, implying the summands $\langle\position_{i}^{\prime}\spin_{i}^{\prime}|(\evolve^{\dagger})^{t}\hat{\position}_{i}^{2}\evolve^{t}|\position_{i}\spin_{i}\rangle$
can contribute to the position variance only when the states of the
particles other than the $i$th particle in the bra $\langle\position_{1}^{\prime}\spin_{1}^{\prime}|\cdots\langle x_{n}^{\prime}\spin_{n}^{\prime}|$
and the ket $|\position_{1}\spin_{1}\rangle\cdots|x_{n}\spin_{n}\rangle$
match. A similar constraint applies to the summands $\langle\position_{j}^{\prime}\spin_{j}^{\prime}|(\evolve^{\dagger})^{t}\hat{\position}_{j}\evolve^{t}|\position_{j}\spin_{j}\rangle\langle\position_{k}^{\prime}\spin_{k}^{\prime}|(\evolve^{\dagger})^{t}\hat{\position}_{k}\evolve^{t}|\position_{k}\spin_{k}\rangle$.
For the single-particle term, the matrix elements can be further derived
through the following procedure,
\begin{equation}
\begin{aligned}\langle\position_{i}^{\prime}\spin_{i}^{\prime}|(\evolve^{\dagger})^{t}\hat{\position}_{i}^{2}\evolve^{t}|\position_{i}\spin_{i}\rangle= & \int_{-\pi}^{\pi}\int_{-\pi}^{\pi}{\rm d}\K^{\prime}{\rm d}\K\sum_{\dspin,\dspin^{\prime}}\langle\position_{i}^{\prime}\spin_{i}^{\prime}|\K^{\prime}\dspin^{\prime}\rangle\langle\K^{\prime}\dspin^{\prime}|(\evolve^{\dagger})^{t}\hat{\position}_{i}^{2}\evolve^{t}|\K\dspin\rangle\langle\K\dspin|\position_{i}\spin_{i}\rangle\\
= & \int_{-\pi}^{\pi}\int_{-\pi}^{\pi}{\rm d}\K^{\prime}{\rm d}\K\sum_{\dspin,\dspin^{\prime}}\langle\position_{i}^{\prime}|\K^{\prime}\rangle\langle\spin_{i}^{\prime}|\dspin^{\prime}\rangle\langle\K|\position_{i}\rangle\langle\dspin|\spin_{i}\rangle\langle\dspin^{\prime}|(\evolve_{K^{\prime}}^{\dagger})^{t}|\dspin^{\prime}\rangle\langle\K^{\prime}|\hat{\position}_{i}^{2}|\K\rangle\langle\dspin^{\prime}|\dspin\rangle\langle\dspin|\evolve_{K}^{t}|\dspin\rangle\\
= & -\frac{1}{2\pi}\int_{-\pi}^{\pi}\int_{-\pi}^{\pi}{\rm d}\K^{\prime}{\rm d}\K\sum_{\dspin,\dspin^{\prime}}e^{i\K^{\prime}x_{i}^{\prime}}e^{-i\K x_{i}}\langle\spin_{i}^{\prime}|\dspin^{\prime}\rangle\langle\dspin|\spin_{i}\rangle\langle\dspin^{\prime}|(\evolve_{K^{\prime}}^{\dagger})^{t}|\dspin^{\prime}\rangle\delta^{(2)}(\K^{\prime}-\K)\langle\dspin^{\prime}|\dspin\rangle\langle\dspin|\evolve_{K}^{t}|\dspin\rangle\\
= & -\frac{1}{2\pi}\int_{-\pi}^{\pi}{\rm d}\K\sum_{\dspin,\dspin^{\prime}}\partial_{\K^{\prime}}^{2}[e^{i(\K^{\prime}x_{i}^{\prime}-\K x_{i})}\langle\spin_{i}^{\prime}|\dspin^{\prime}\rangle\langle\dspin|\spin_{i}\rangle\lambda_{\dspin^{\prime}}^{*}(\K^{\prime})^{t}\lambda_{d}(\K)^{t}\langle\dspin^{\prime}(\K^{\prime})|\dspin(\K)\rangle]|_{\K^{\prime}=\K}.
\end{aligned}
\end{equation}
We used
\begin{equation}
\begin{aligned}\langle\K^{\prime}|\hat{\position}_{i}^{2}|\K\rangle= & \sum_{\position=-\infty}^{\infty}\langle\K^{\prime}|\hat{\position}_{i}^{2}|\position\rangle\langle\position|\K\rangle=\frac{1}{2\pi}\sum_{x=-\infty}^{\infty}x^{2}e^{-i(\K^{\prime}-\K)x}=-\sum_{l=-\infty}^{\infty}\delta^{(2)}(\K^{\prime}-\K+2\pi l),\end{aligned}
\end{equation}
$l\in\mathbb{Z}$, where we utilized the relation
\begin{equation}
\frac{1}{2\pi}\sum_{\position=-\infty}^{\infty}\position^{m}e^{-i(\K^{\prime}-\K)\position}=(i)^{m}\sum_{l=-\infty}^{\infty}\delta^{(m)}(\K^{\prime}-\K+2\pi l),
\end{equation}
which is the periodic extension of the derivative of the Dirac delta
function with period $2\pi$. We can omit the summation on the right-hand
side because the integration region is $2\pi$, so there is one term
remaining. Then we can use $\langle\K^{\prime}|\hat{\position}_{i}^{2}|\K\rangle=-\delta^{(2)}(\K^{\prime}-\K)$
in the integral.

The order of $t$ is relative to the order of derivatives. If we expand
the derivatives,
\begin{equation}
\begin{aligned} & \langle\position_{i}^{\prime}\spin_{i}^{\prime}|(\evolve^{\dagger})^{t}\hat{\position}_{i}^{2}\evolve^{t}|\position_{i}\spin_{i}\rangle\\
= & -\frac{1}{2\pi}\int_{-\pi}^{\pi}{\rm d}\K\sum_{\dspin,\dspin^{\prime}}\partial_{\K^{\prime}}^{2}[e^{i\K^{\prime}x_{i}^{\prime}}\langle\spin_{i}^{\prime}|\dspin^{\prime}\rangle\lambda_{\dspin^{\prime}}^{*}(\K^{\prime})^{t}\langle\dspin^{\prime}(\K^{\prime})|\dspin(\K)\rangle]|_{K^{\prime}=\K}e^{-i\K x_{i}}\langle\dspin|\spin_{i}\rangle\lambda_{d}(\K)^{t}\\
= & -\frac{1}{2\pi}\int_{-\pi}^{\pi}{\rm d}\K\sum_{\dspin,\dspin^{\prime}}\{\partial_{\K^{\prime}}^{2}[e^{i\K^{\prime}x_{i}^{\prime}}\langle\spin_{i}^{\prime}|\dspin^{\prime}\rangle\langle\dspin^{\prime}(\K^{\prime})|\dspin(\K)\rangle]\lambda_{\dspin^{\prime}}^{*}(\K^{\prime})^{t}+2\partial_{\K^{\prime}}[e^{i\K^{\prime}x_{i}^{\prime}}\langle\spin_{i}^{\prime}|\dspin^{\prime}\rangle\langle\dspin^{\prime}(\K^{\prime})|\dspin(\K)\rangle]\partial_{\K^{\prime}}[\lambda_{\dspin^{\prime}}^{*}(\K^{\prime})^{t}]\\
 & +e^{i\K^{\prime}x_{i}^{\prime}}\langle\spin_{i}^{\prime}|\dspin^{\prime}\rangle\langle\dspin^{\prime}(\K^{\prime})|\dspin(\K)\rangle\partial_{\K^{\prime}}^{2}[\lambda_{\dspin^{\prime}}^{*}(\K^{\prime})^{t}]\}|_{\K^{\prime}=\K}e^{-i\K x_{i}}\langle\dspin|\spin_{i}\rangle\lambda_{d}(\K)^{t}\\
= & -\frac{1}{2\pi}\int_{-\pi}^{\pi}{\rm d}\K\sum_{\dspin}\{\partial_{\K^{\prime}}^{2}[e^{i\K^{\prime}x_{i}^{\prime}}\langle\spin_{i}^{\prime}|\dspin(\K^{\prime})\rangle\langle\dspin(\K^{\prime})|\dspin(\K)\rangle]\lambda_{\dspin}^{*}(\K^{\prime})^{t}+2\partial_{\K^{\prime}}[e^{i\K^{\prime}x_{i}^{\prime}}\langle\spin_{i}^{\prime}|\dspin(\K^{\prime})\rangle\langle\dspin(\K^{\prime})|\dspin(\K)\rangle]\partial_{\K^{\prime}}[\lambda_{\dspin}^{*}(\K^{\prime})^{t}]\\
 & +e^{i\K^{\prime}x_{i}^{\prime}}\langle\spin_{i}^{\prime}|\dspin(\K^{\prime})\rangle\langle\dspin(\K^{\prime})|\dspin(\K)\rangle\partial_{\K^{\prime}}^{2}[\lambda_{\dspin}^{*}(\K^{\prime})^{t}]\}|_{K^{\prime}=\K}e^{-i\K x_{i}}\langle\dspin(\K)|\spin_{i}\rangle\lambda_{d}(\K)^{t}\\
 & -\frac{1}{2\pi}\int_{-\pi}^{\pi}{\rm d}\K\sum_{\dspin\ne\dspin^{\prime}}\{\partial_{\K^{\prime}}^{2}[e^{i\K^{\prime}x_{i}^{\prime}}\langle\spin_{i}^{\prime}|\dspin^{\prime}(\K^{\prime})\rangle\langle\dspin^{\prime}(\K^{\prime})|\dspin(\K)\rangle]\lambda_{\dspin^{\prime}}^{*}(\K^{\prime})^{t}+2\partial_{\K^{\prime}}[e^{i\K^{\prime}x_{i}^{\prime}}\langle\spin_{i}^{\prime}|\dspin^{\prime}(\K^{\prime})\rangle\langle\dspin^{\prime}(\K^{\prime})|\dspin(\K)\rangle]\partial_{\K^{\prime}}[\lambda_{\dspin^{\prime}}^{*}(\K^{\prime})^{t}]\\
 & +e^{i\K^{\prime}x_{i}^{\prime}}\langle\spin_{i}^{\prime}|\dspin^{\prime}(\K^{\prime})\rangle\langle\dspin^{\prime}(\K^{\prime})|\dspin(\K)\rangle\partial_{\K^{\prime}}^{2}[\lambda_{\dspin^{\prime}}^{*}(\K^{\prime})^{t}]\}|_{\K^{\prime}=\K}e^{-i\K x_{i}}\langle\dspin(\K)|\spin_{i}\rangle\lambda_{d}(\K)^{t}\\
= & -\frac{1}{2\pi}\int_{-\pi}^{\pi}{\rm d}\K\sum_{\dspin}\{\partial_{\K^{\prime}}^{2}[e^{i\K^{\prime}x_{i}^{\prime}}\langle\spin_{i}^{\prime}|\dspin(\K^{\prime})\rangle\langle\dspin(\K^{\prime})|\dspin(\K)\rangle]|_{\K^{\prime}=\K}e^{-i\K x_{i}}\langle\dspin(\K)|\spin_{i}\rangle\\
 & +2\partial_{\K^{\prime}}[e^{i\K^{\prime}x_{i}^{\prime}}\langle\spin_{i}^{\prime}|\dspin(\K^{\prime})\rangle\langle\dspin(\K^{\prime})|\dspin(\K)\rangle]|_{\K^{\prime}=\K}t\lambda_{\dspin}^{*}(\K)^{-1}\partial_{\K}[\lambda_{\dspin}^{*}(\K)]e^{-i\K x_{i}}\langle\dspin(\K)|\spin_{i}\rangle\\
 & +e^{i\K(x_{i}^{\prime}-x_{i})}\langle\spin_{i}^{\prime}|\dspin(\K)\rangle t(t-1)\lambda_{\dspin}^{*}(\K)^{-2}\langle\dspin(\K)|\spin_{i}\rangle(\partial_{\K}[\lambda_{\dspin}^{*}(\K)])^{2}+e^{i\K(x_{i}^{\prime}-x_{i})}\langle\spin_{i}^{\prime}|\dspin(\K)\rangle\langle\dspin(\K)|\spin_{i}\rangle t\lambda_{\dspin}^{*}(\K)^{-1}\partial_{\K}^{2}[\lambda_{\dspin}^{*}(\K)]\}\}\\
 & -\frac{1}{2\pi}\int_{-\pi}^{\pi}{\rm d}\K\sum_{\dspin\ne\dspin^{\prime}}\{\partial_{\K^{\prime}}^{2}[e^{i\K^{\prime}x_{i}^{\prime}}\langle\spin_{i}^{\prime}|\dspin^{\prime}(\K^{\prime})\rangle\langle\dspin^{\prime}(\K^{\prime})|\dspin(\K)\rangle]|_{K^{\prime}=\K}\\
 & +2\partial_{\K^{\prime}}[e^{i\K^{\prime}x_{i}^{\prime}}\langle\spin_{i}^{\prime}|\dspin^{\prime}(\K^{\prime})\rangle\langle\dspin^{\prime}(\K^{\prime})|\dspin(\K)\rangle]|_{K^{\prime}=\K}t\lambda_{\dspin^{\prime}}^{*}(\K)^{-1}\partial_{\K}[\lambda_{\dspin^{\prime}}^{*}(\K)]\\
 & +e^{i\K x_{i}^{\prime}}\langle\spin_{i}^{\prime}|\dspin^{\prime}(\K)\rangle\langle\dspin^{\prime}(\K)|\dspin(\K)\rangle\{t(t-1)\lambda_{\dspin^{\prime}}^{*}(\K)^{-2}(\partial_{\K}[\lambda_{\dspin^{\prime}}^{*}(\K)])^{2}+t\lambda_{\dspin^{\prime}}^{*}(\K)^{-1}\partial_{\K}^{2}[\lambda_{\dspin^{\prime}}^{*}(\K)]\}\}e^{-i\K x_{i}}\langle\dspin(\K)|\spin_{i}\rangle\lambda_{\dspin^{\prime}}^{*}(\K)^{t}\lambda_{d}(\K)^{t}\\
= & -\frac{1}{2\pi}[t^{2}\IA(\position_{i}^{\prime}-\position_{i},\spin_{i}^{\prime},\spin_{i})+t\IAM(\position_{i}^{\prime},\position_{i},\spin_{i}^{\prime},\spin_{i})+\IAC(\position_{i}^{\prime},\position_{i},\spin_{i}^{\prime},\spin_{i})+\IAoscill(\position_{i}^{\prime},\position_{i},\spin_{i}^{\prime},\spin_{i};t)].
\end{aligned}
\end{equation}
Denote
\begin{equation}
\IA(\position_{i}^{\prime}-\position_{i},\spin_{i}^{\prime},\spin_{i})=\int_{-\pi}^{\pi}{\rm d}\K e^{i\K(x_{i}^{\prime}-x_{i})}\sum_{\dspin}\langle\spin_{i}^{\prime}|\dspin(\K)\rangle\langle\dspin(\K)|\spin_{i}\rangle\lambda_{\dspin}^{*}(\K)^{-2}(\partial_{\K}[\lambda_{\dspin}^{*}(\K)])^{2},
\end{equation}
\begin{equation}
\begin{aligned}\IAM(\position_{i}^{\prime},\position_{i},\spin_{i}^{\prime},\spin_{i})= & \int_{-\pi}^{\pi}{\rm d}\K\sum_{\dspin}\{2\partial_{\K^{\prime}}[e^{i\K^{\prime}x_{i}^{\prime}}\langle\spin_{i}^{\prime}|\dspin(\K^{\prime})\rangle\langle\dspin(\K^{\prime})|\dspin(\K)\rangle]|_{\K^{\prime}=\K}\lambda_{\dspin}^{*}(\K)^{-1}\partial_{\K}[\lambda_{\dspin}^{*}(\K)]e^{-i\K x_{i}}\langle\dspin(\K)|\spin_{i}\rangle\\
 & +e^{i\K(x_{i}^{\prime}-x_{i})}\langle\spin_{i}^{\prime}|\dspin(\K)\rangle\langle\dspin(\K)|\spin_{i}\rangle\{-\lambda_{\dspin}^{*}(\K)^{-2}(\partial_{\K}[\lambda_{\dspin}^{*}(\K)])^{2}+\lambda_{\dspin}^{*}(\K)^{-1}\partial_{\K}^{2}[\lambda_{\dspin}^{*}(\K)]\}\},
\end{aligned}
\end{equation}
\begin{equation}
\begin{aligned}\IAC(\position_{i}^{\prime},\position_{i},\spin_{i}^{\prime},\spin_{i})= & \int_{-\pi}^{\pi}{\rm d}\K\sum_{\dspin}\partial_{\K^{\prime}}^{2}[e^{i\K^{\prime}x_{i}^{\prime}}\langle\spin_{i}^{\prime}|\dspin(\K^{\prime})\rangle\langle\dspin(\K^{\prime})|\dspin(\K)\rangle]|_{\K^{\prime}=\K}e^{-i\K x_{i}}\langle\dspin(\K)|\spin_{i}\rangle,\end{aligned}
\end{equation}
\begin{equation}
\begin{aligned}\IAoscill(\position_{i}^{\prime},\position_{i},\spin_{i}^{\prime},\spin_{i};t)= & \int_{-\pi}^{\pi}{\rm d}\K\sum_{\dspin\ne\dspin^{\prime}}\{\partial_{\K^{\prime}}^{2}[e^{i\K^{\prime}x_{i}^{\prime}}\langle\spin_{i}^{\prime}|\dspin^{\prime}(\K^{\prime})\rangle\langle\dspin^{\prime}(\K^{\prime})|\dspin(\K)\rangle]|_{\K^{\prime}=\K}\\
 & +2\partial_{\K^{\prime}}[e^{i\K^{\prime}x_{i}^{\prime}}\langle\spin_{i}^{\prime}|\dspin^{\prime}(\K^{\prime})\rangle\langle\dspin^{\prime}(\K^{\prime})|\dspin(\K)\rangle]|_{\K^{\prime}=\K}t\lambda_{\dspin^{\prime}}^{*}(\K)^{-1}\partial_{\K}[\lambda_{\dspin^{\prime}}^{*}(\K)]\}e^{-i\K x_{i}}\langle\dspin(\K)|\spin_{i}\rangle(\lambda_{\dspin^{\prime}}^{*}(\K)\lambda_{d}(\K))^{t}.
\end{aligned}
\end{equation}
And for the other term,
\begin{equation}
\begin{aligned} & \langle\position_{i}^{\prime}\spin_{i}^{\prime}|(\evolve^{\dagger})^{t}\hat{\position}_{i}\evolve^{t}|\position_{i}\spin_{i}\rangle\\
= & \int_{-\pi}^{\pi}\int_{-\pi}^{\pi}{\rm d}\K^{\prime}{\rm d}\K\sum_{\dspin,\dspin^{\prime}}\langle\position_{i}^{\prime}\spin_{i}^{\prime}|\K^{\prime}\dspin^{\prime}\rangle\langle\K^{\prime}\dspin^{\prime}|(\evolve^{\dagger})^{t}\hat{\position}_{i}\evolve^{t}|\K\dspin\rangle\langle\K\dspin|\position_{i}\spin_{i}\rangle\\
= & \int_{-\pi}^{\pi}\int_{-\pi}^{\pi}{\rm d}\K^{\prime}{\rm d}\K\sum_{\dspin,\dspin^{\prime}}\langle\position_{i}^{\prime}|\K^{\prime}\rangle\langle\spin_{i}^{\prime}|\dspin^{\prime}\rangle\langle\K|\position_{i}\rangle\langle\dspin|\spin_{i}\rangle\langle\dspin^{\prime}|(\evolve_{K^{\prime}}^{\dagger})^{t}|\dspin^{\prime}\rangle\langle\K^{\prime}|\hat{\position}_{i}|\K\rangle\langle\dspin^{\prime}|\dspin\rangle\langle\dspin|\evolve_{K}^{t}|\dspin\rangle\\
= & \frac{i}{2\pi}\int_{-\pi}^{\pi}\int_{-\pi}^{\pi}{\rm d}\K^{\prime}{\rm d}\K\sum_{\dspin,\dspin^{\prime}}e^{i\K^{\prime}x_{i}^{\prime}}e^{-i\K x_{i}}\langle\spin_{i}^{\prime}|\dspin^{\prime}\rangle\langle\dspin|\spin_{i}\rangle\langle\dspin^{\prime}|(\evolve_{K^{\prime}}^{\dagger})^{t}|\dspin^{\prime}\rangle\delta^{(1)}(\K^{\prime}-\K)\langle\dspin^{\prime}|\dspin\rangle\langle\dspin|\evolve_{K}^{t}|\dspin\rangle\\
= & -\frac{i}{2\pi}\int_{-\pi}^{\pi}{\rm d}\K\sum_{\dspin,\dspin^{\prime}}\partial_{\K^{\prime}}[e^{i(\K^{\prime}x_{i}^{\prime}-\K x_{i})}\langle\spin_{i}^{\prime}|\dspin^{\prime}\rangle\langle\dspin|\spin_{i}\rangle\lambda_{\dspin^{\prime}}^{*}(\K^{\prime})^{t}\lambda_{d}(\K)^{t}\langle\dspin^{\prime}(\K^{\prime})|\dspin(\K)\rangle]|_{\K^{\prime}=\K}.
\end{aligned}
\end{equation}
Similarly,
\begin{equation}
\begin{aligned}\langle\position_{i}^{\prime}\spin_{i}^{\prime}|(\evolve^{\dagger})^{t}\hat{\position}_{i}\evolve^{t}|\position_{i}\spin_{i}\rangle= & -\frac{i}{2\pi}\int_{-\pi}^{\pi}{\rm d}\K\sum_{\dspin,\dspin^{\prime}}\{\partial_{\K^{\prime}}[e^{i\K^{\prime}x_{i}^{\prime}}\langle\spin_{i}^{\prime}|\dspin^{\prime}(\K^{\prime})\rangle\langle\dspin^{\prime}(\K^{\prime})|\dspin(\K)\rangle]\lambda_{\dspin^{\prime}}^{*}(\K^{\prime})^{t}\\
 & \ +e^{i\K^{\prime}x_{i}^{\prime}}\langle\spin_{i}^{\prime}|\dspin^{\prime}(\K^{\prime})\rangle\langle\dspin^{\prime}(\K^{\prime})|\dspin(\K)\rangle\partial_{\K^{\prime}}[\lambda_{\dspin^{\prime}}^{*}(\K^{\prime})^{t}]\}|_{\K^{\prime}=\K}e^{-i\K x_{i}}\langle\dspin(\K)|\spin_{i}\rangle\lambda_{d}(\K)^{t}\\
= & -\frac{i}{2\pi}\int_{-\pi}^{\pi}{\rm d}\K\sum_{\dspin}\{\partial_{\K^{\prime}}[e^{i\K^{\prime}x_{i}^{\prime}}\langle\spin_{i}^{\prime}|\dspin(\K^{\prime})\rangle\langle\dspin(\K^{\prime})|\dspin(\K)\rangle]\lambda_{\dspin}^{*}(\K^{\prime})^{t}\\
 & \ +e^{i\K^{\prime}x_{i}^{\prime}}\langle\spin_{i}^{\prime}|\dspin(\K^{\prime})\rangle\langle\dspin(\K^{\prime})|\dspin(\K)\rangle\partial_{\K^{\prime}}[\lambda_{\dspin}^{*}(\K^{\prime})^{t}]\}|_{K^{\prime}=\K}e^{-i\K x_{i}}\langle\dspin(\K)|\spin_{i}\rangle\lambda_{d}(\K)^{t}\\
 & -\frac{i}{2\pi}\int_{-\pi}^{\pi}{\rm d}\K\sum_{\dspin^{\prime}\ne\dspin}\{\partial_{\K^{\prime}}[e^{i\K^{\prime}x_{i}^{\prime}}\langle\spin_{i}^{\prime}|\dspin^{\prime}(\K^{\prime})\rangle\langle\dspin^{\prime}(\K^{\prime})|\dspin(\K)\rangle]|_{K^{\prime}=\K}\lambda_{\dspin^{\prime}}^{*}(\K)^{t}\\
 & \ +e^{i\K x_{i}^{\prime}}\langle\spin_{i}^{\prime}|\dspin^{\prime}(\K)\rangle\langle\dspin^{\prime}(\K)|\dspin(\K)\rangle\partial_{\K^{\prime}}[\lambda_{\dspin^{\prime}}^{*}(\K^{\prime})^{t}]|_{K^{\prime}=\K}\}e^{-i\K x_{i}}\langle\dspin(\K)|\spin_{i}\rangle\lambda_{d}(\K)^{t}\\
= & -\frac{i}{2\pi}\int_{-\pi}^{\pi}{\rm d}\K\sum_{\dspin}\{\partial_{\K^{\prime}}[e^{i\K^{\prime}x_{i}^{\prime}}\langle\spin_{i}^{\prime}|\dspin(\K^{\prime})\rangle\langle\dspin(\K^{\prime})|\dspin(\K)\rangle]|_{\K^{\prime}=\K}\lambda_{\dspin}^{*}(\K)^{t}\\
 & \ +e^{i\K x_{i}^{\prime}}\langle\spin_{i}^{\prime}|\dspin(\K)\rangle\partial_{\K^{\prime}}[\lambda_{\dspin}^{*}(\K^{\prime})^{t}]|_{\K^{\prime}=\K}\}e^{-i\K x_{i}}\langle\dspin(\K)|\spin_{i}\rangle\lambda_{d}(\K)^{t}\\
 & -\frac{i}{2\pi}\int_{-\pi}^{\pi}{\rm d}\K\sum_{\dspin^{\prime}\ne\dspin}\partial_{\K^{\prime}}[e^{i\K^{\prime}x_{i}^{\prime}}\langle\spin_{i}^{\prime}|\dspin^{\prime}(\K^{\prime})\rangle\langle\dspin^{\prime}(\K^{\prime})|\dspin(\K)\rangle]|_{\K^{\prime}=\K}e^{-iKx_{i}}\langle\dspin(\K)|\spin_{i}\rangle(\lambda_{\dspin^{\prime}}^{*}(\K)\lambda_{d}(\K))^{t}\\
= & -\frac{i}{2\pi}[t\IB(\position_{i}^{\prime}-\position_{i},\spin_{i}^{\prime},\spin_{i})+\IBM(\position_{i}^{\prime},\position_{i},\spin_{i}^{\prime},\spin_{i})+\IBoscill(\position_{i}^{\prime},\position_{i},\spin_{i}^{\prime},\spin_{i};t)],
\end{aligned}
\end{equation}
\begin{equation}
\IB(\position_{i}^{\prime}-\position_{i},\spin_{i}^{\prime},\spin_{i})=\int_{-\pi}^{\pi}{\rm d}\K\sum_{\dspin}e^{i\K(x_{i}^{\prime}-x_{i})}\langle\spin_{i}^{\prime}|\dspin(\K)\rangle\langle\dspin(\K)|\spin_{i}\rangle\lambda_{\dspin}^{*}(\K)^{-1}\partial_{\K}[\lambda_{\dspin}^{*}(\K)],
\end{equation}
\begin{equation}
\begin{aligned}\IBM(\position_{i}^{\prime},\position_{i},\spin_{i}^{\prime},\spin_{i})= & \int_{-\pi}^{\pi}{\rm d}\K\sum_{\dspin}\partial_{\K^{\prime}}[e^{i\K^{\prime}x_{i}^{\prime}}\langle\spin_{i}^{\prime}|\dspin(\K^{\prime})\rangle\langle\dspin(\K^{\prime})|\dspin(\K)\rangle]|_{\K^{\prime}=\K}e^{-i\K x_{i}}\langle\dspin(\K)|\spin_{i}\rangle\\
= & \int_{-\pi}^{\pi}{\rm d}\K\sum_{\dspin}\{\partial_{\K^{\prime}}[e^{i\K^{\prime}x_{i}^{\prime}}\langle\spin_{i}^{\prime}|\dspin(\K^{\prime})\rangle]|_{\K^{\prime}=\K}+e^{i\K x_{i}^{\prime}}\langle\spin_{i}^{\prime}|\dspin(\K)\rangle\partial_{\K^{\prime}}[\langle\dspin(\K^{\prime})|\dspin(\K)\rangle]|_{\K^{\prime}=\K}\}e^{-i\K x_{i}}\langle\dspin(\K)|\spin_{i}\rangle,
\end{aligned}
\end{equation}
\begin{equation}
\begin{aligned}\IBoscill(\position_{i}^{\prime},\position_{i},\spin_{i}^{\prime},\spin_{i};t)= & \int_{-\pi}^{\pi}{\rm d}\K\sum_{\dspin\ne\dspin^{\prime}}\partial_{\K^{\prime}}[e^{i\K^{\prime}x_{i}^{\prime}}\langle\spin_{i}^{\prime}|\dspin^{\prime}(\K^{\prime})\rangle\langle\dspin^{\prime}(\K^{\prime})|\dspin(\K)\rangle]|_{\K^{\prime}=\K}e^{-i\K x_{i}}\langle\dspin(\K)|\spin_{i}\rangle(\lambda_{\dspin^{\prime}}^{*}(\K)\lambda_{d}(\K))^{t}\\
= & \int_{-\pi}^{\pi}{\rm d}\K\sum_{\dspin\ne\dspin^{\prime}}e^{i\K(x_{i}^{\prime}-x_{i})}\langle\spin_{i}^{\prime}|\dspin^{\prime}(\K)\rangle\langle\dspin(\K)|\spin_{i}\rangle\partial_{\K^{\prime}}[\langle\dspin^{\prime}(\K^{\prime})|\dspin(\K)\rangle]|_{\K^{\prime}=\K}(\lambda_{\dspin^{\prime}}^{*}(\K)\lambda_{d}(\K))^{t}.
\end{aligned}
\end{equation}

The above expressions apply to any coin operator of SU(2) with $\lambda_{d}(K)$
and $|d(K)\rangle$ computed from the conditional unitary operator
$\hat{U}_{K}$ with the new coin operator. If we use the Hadamard
coin,
\begin{equation}
\begin{aligned}\IA(\position^{\prime}-\position,\spin^{\prime},\spin)= & \begin{cases}
(\sqrt{2}-2)\pi, & \position^{\prime}-\position=0,\spin^{\prime}=\spin\\
\sqrt{2}\pi\left(\sqrt{2}-1\right)^{|\position^{\prime}-\position|}\cos\left(\frac{\pi(\position^{\prime}-\position)}{2}\right), & \position^{\prime}-\position\ne0,\spin^{\prime}=\spin\\
0, & \spin^{\prime}\ne\spin
\end{cases}\\
= & \begin{cases}
(1-\sqrt{2})f(0), & \position^{\prime}-\position=0,\spin^{\prime}=\spin\\
\f(\position^{\prime}-\position), & \position^{\prime}-\position\ne0,\spin^{\prime}=\spin\\
0, & \spin^{\prime}\ne\spin,
\end{cases}
\end{aligned}
\end{equation}
where $\f(x)=\sqrt{2}\pi\left(\sqrt{2}-1\right)^{|\position|}\cos\left(\pi x/2\right)$.
\begin{equation}
\IAM(\position^{\prime},\position,\spin^{\prime},\spin)=\begin{cases}
\begin{aligned}16\position^{\prime}[\IAMP(\Delta\position-2)+\IAMP(\Delta\position+2)]+28\position^{\prime}\IAMP(\Delta\position)\\
+2\position^{\prime}[\IAMP(\Delta\position+4)+\IAMP(\Delta\position-4)] & +8[\IAMP(\Delta\position+2)-\IAMP(\Delta\position-2)],
\end{aligned}
 & \spin^{\prime}=\spin=\uparrow\\
\begin{aligned}-\{16\position^{\prime}[\IAMP(\Delta\position-2)+\IAMP(\Delta\position+2)]+28\position^{\prime}\IAMP(\Delta\position)\\
+2\position^{\prime}[\IAMP(\Delta\position+4)+\IAMP(\Delta\position-4)] & +8[\IAMP(\Delta\position+2)-\IAMP(\Delta\position-2)]\},
\end{aligned}
 & s^{\prime}=\spin=\downarrow\\
\begin{aligned}(4x^{\prime}-4)\IAMP(\Delta\position+2)+(28\position^{\prime}-8)\IAMP(\Delta\position)+(28x^{\prime}-20)\IAMP(\Delta\position-2)+4\position^{\prime}\IAMP(\Delta\position-4),\end{aligned}
 & \spin^{\prime}=\uparrow,\spin=\downarrow\\
\begin{aligned}(4x^{\prime}+4)\IAMP(\Delta\position-2)+(28\position^{\prime}+8)\IAMP(\Delta\position) & +(28x^{\prime}+20)\IAMP(\Delta\position+2)+4\position^{\prime}\IAMP(\Delta\position+4),\end{aligned}
 & \spin^{\prime}=\downarrow,\spin=\uparrow,
\end{cases}
\end{equation}
\begin{equation}
\IAC(\position^{\prime},\position,\spin^{\prime},\spin)=\begin{cases}
\begin{aligned}\position^{\prime2}\IAMP(\position^{\prime}-\position-4)-4(1+\position^{\prime}-3\position^{\prime2})\IAMP(\position^{\prime}-\position-2)+(16-24\position^{\prime}+38\position^{\prime2})\IAMP(\position^{\prime}-\position)\\
+4(1-\position^{\prime}+3\position^{\prime2})\IAMP(\position^{\prime}-\position+2)+\position^{\prime2}\IAMP(\position^{\prime}-\position+4),
\end{aligned}
 & \spin^{\prime}=\spin=\uparrow;\\
\begin{aligned}\position^{\prime2}\IAMP(\position^{\prime}-\position-4)+4(1+\position^{\prime}+3\position^{\prime2})\IAMP(\position^{\prime}-\position-2)+(16+24\position^{\prime}+38\position^{\prime2})\IAMP(\position^{\prime}-\position)\\
+4(-1+\position^{\prime}+3\position^{\prime2})\IAMP(\position^{\prime}-\position+2)+\position^{\prime2}\IAMP(\position^{\prime}-\position+4),
\end{aligned}
 & \spin^{\prime}=\spin=\downarrow;\\
4(1+\position^{\prime})\IAMP(\position^{\prime}-\position-2)+24\position^{\prime}\IAMP(\position^{\prime}-\position)+4(\position^{\prime}-1)\IAMP(\position^{\prime}-\position+2), & \spin^{\prime}\ne\spin,
\end{cases}
\end{equation}
where $\Delta\position=\position^{\prime}-\position$, $\IAMP(x)=-\frac{\pi}{64}(\sqrt{2}-1)^{|x|}(3\sqrt{2}+2|x|)\cos(\frac{\pi x}{2})$.
\begin{equation}
\begin{aligned} & \IAoscill(\position^{\prime},\position,\spin^{\prime},\spin;t)\\
= & \begin{cases}
(-1)^{\frac{3}{4}}\sqrt{\frac{\pi}{4t}}e^{-\frac{1}{2}i\pi(3t+x^{\prime}+x)}\left((-1+(1+i)x^{\prime})(e^{i\pi(t+x)}+e^{i\pi(3t+\position^{\prime})})+((1+i)x^{\prime}-i)(e^{i\pi(2t+x)}+e^{i\pi x^{\prime}})\right), & \spin^{\prime}=\spin=\uparrow\\
-(-1)^{\frac{3}{4}}\sqrt{\frac{\pi}{4t}}e^{-\frac{1}{2}i\pi(3t+x^{\prime}+x)}\left((1+(1+i)x^{\prime})(e^{i\pi(t+x)}+e^{i\pi(3t+x^{\prime})})+((1+i)x^{\prime}+i)(e^{i\pi(2t+x)}+e^{i\pi x^{\prime}})\right), & \spin^{\prime}=\spin=\downarrow\\
(-1)^{\frac{1}{4}}\sqrt{\frac{\pi}{4t}}e^{-\frac{1}{2}i\pi(3t+x^{\prime}+x)}\left(((1+i)x^{\prime}-i)(e^{i\pi(2t+x)}+e^{i\pi x^{\prime}})+(1-(1+i)x^{\prime})(e^{i\pi(t+x)}+e^{i\pi(3t+x^{\prime})})\right), & \spin^{\prime}=\uparrow,\spin=\downarrow\\
(-1)^{\frac{3}{4}}\sqrt{\frac{\pi}{4t}}e^{-\frac{1}{2}i\pi(3t+\position^{\prime}+\position)}\left((1+(1-i)x^{\prime})(e^{i\pi(2t+x)}+e^{i\pi x^{\prime}})+i(1+(1+i)x^{\prime})(e^{i\pi(t+x)}+e^{i\pi(3t+x^{\prime})})\right), & \spin^{\prime}=\downarrow,\spin=\uparrow
\end{cases}\\
 & +o(\sqrt{\frac{1}{t}}).
\end{aligned}
\end{equation}
\begin{equation}
\begin{aligned}\IB(\position^{\prime}-\position,\spin^{\prime},\spin)= & \begin{cases}
-i\IA(\position^{\prime}-\position,\spin^{\prime},\spin), & \spin^{\prime}=s=\uparrow\\
i\IA(\position^{\prime}-\position,\spin^{\prime},\spin), & \spin^{\prime}=\spin=\downarrow\\
\frac{i\pi}{\sqrt{2}}\cos(\frac{\pi(\position^{\prime}-\position)}{2})[(\sqrt{2}-1)^{|\position^{\prime}-\position|}-(\sqrt{2}-1)^{|\position^{\prime}-\position-2|}], & \spin^{\prime}=\uparrow,\spin=\downarrow\\
\frac{i\pi}{\sqrt{2}}\cos(\frac{\pi(\position^{\prime}-\position)}{2})[(\sqrt{2}-1)^{|\position^{\prime}-\position|}-(\sqrt{2}-1)^{|\position^{\prime}-\position+2|}], & \spin^{\prime}=\downarrow,\spin=\uparrow
\end{cases}\\
= & \begin{cases}
(-1)^{\spin}i\IA(\position^{\prime}-\position,\spin^{\prime},\spin), & \spin^{\prime}=s\\
\frac{i}{2}[f(\position^{\prime}-\position)+f(\position^{\prime}-\position-(-1)^{\spin}\times2)], & \spin^{\prime}\ne\spin.
\end{cases}
\end{aligned}
\end{equation}
In the above equation, when $\spin$ is $\uparrow$, it acts as number
$1$, and $\downarrow$ as number $0$; i.e., when $\spin=\uparrow$,
$(-1)^{\spin}=-1$, and when $\spin=\downarrow$, $(-1)^{\spin}=1$.
\begin{equation}
\begin{aligned}\IBM(\position^{\prime},\position,\spin^{\prime},\spin)= & \begin{cases}
2i\pi x\delta_{x^{\prime}-x,0}-\frac{i\pi\left(\sqrt{2}-1\right)^{|x^{\prime}-x|}\cos\left(\frac{\pi(x^{\prime}-x)}{2}\right)}{\sqrt{2}}, & \spin^{\prime}=\spin=\uparrow\\
2i\pi x\delta_{x^{\prime}-x,0}+\frac{i\pi\left(\sqrt{2}-1\right)^{|x^{\prime}-x|}\cos\left(\frac{\pi(x^{\prime}-x)}{2}\right)}{\sqrt{2}}, & \spin^{\prime}=\spin=\downarrow\\
\frac{i\pi\left(\sqrt{2}-1\right)^{|x^{\prime}-x|}\cos\left(\frac{\pi(x^{\prime}-x)}{2}\right)}{\sqrt{2}}, & \spin^{\prime}\ne\spin.
\end{cases}\end{aligned}
\end{equation}
{\small
\begin{equation}
\begin{aligned}\IBoscill(\position^{\prime},\position,\spin^{\prime},\spin;t)= & \begin{cases}
\frac{1}{2}i\sqrt{\frac{\pi}{2t}}\left(1+e^{i\pi t}\right)e^{-\frac{1}{2}i\pi(3t+x^{\prime}+x)}\left(-e^{i\pi(t+x^{\prime})}+e^{i\pi(2t+x^{\prime})}+e^{i\pi(t+x)}+e^{i\pi x^{\prime}}\right), & s^{\prime}=s=\uparrow\\
-\frac{1}{2}i\sqrt{\frac{\pi}{2t}}\left(1+e^{i\pi t}\right)e^{-\frac{1}{2}i\pi(3t+x^{\prime}+x)}\left(-e^{i\pi(t+x^{\prime})}+e^{i\pi(2t+x^{\prime})}+e^{i\pi(t+x)}+e^{i\pi x^{\prime}}\right), & s^{\prime}=s=\downarrow\\
-\frac{1}{2}\sqrt{\frac{\pi}{2t}}\left(-1+e^{i\pi t}\right)e^{-\frac{1}{2}i\pi(3t+x^{\prime}+x)}\left(e^{i\pi(t+x^{\prime})}+e^{i\pi(2t+x^{\prime})}-e^{i\pi(t+x)}+e^{i\pi x^{\prime}}\right), & s^{\prime}=\uparrow,s=\downarrow\\
-\frac{1}{2}\sqrt{\frac{\pi}{2t}}\left(-1+e^{i\pi t}\right)e^{-\frac{1}{2}i\pi(3t+x^{\prime}+x)}\left(e^{i\pi(t+x^{\prime})}+e^{i\pi(2t+x^{\prime})}-e^{i\pi(t+x)}+e^{i\pi x^{\prime}}\right), & s^{\prime}=\downarrow,s=\uparrow
\end{cases}\end{aligned}
+o(\sqrt{\frac{1}{t}}).
\end{equation}
}$\IAoscill$ and $\IBoscill$ are integrals with rapidly varying
phase when $t$ is large, and they approach zero when $t\rightarrow\infty$.
But note that in the expression of $\langle\distanceOperator\rangle$,
there are terms of $I_{Bo}$ multiplied by $t$, and this results
in the terms of order $\sqrt{t}.$ The asymptotic formula can be calculated
by stationary phase approximation. Let $\Omega$ denote the set of
critical points of a real function $f(x)$ in the region $(a,b)$
and $t>0$, and assume $g(x)$ is compactly supported or has exponential
decay. Then,
\begin{equation}
\int_{a}^{b}g(x)e^{itf(x)}{\rm d}x=\sum_{x_{0}\in\Omega}g(x_{0})e^{itf(x_{0})+\mathrm{sign}(f^{\prime\prime}(x_{0}))\frac{i\pi}{4}}\left(\frac{2\pi}{t|f^{\prime\prime}(x_{0})|}\right)^{\frac{1}{2}}+o(t^{-\frac{1}{2}}).
\end{equation}

To illustrate how we calculate the integrals like $\IA$, $\IAM$,
$\IAC$, $\IB$, and $\IBM$ using the residue theorem, we give an
example of calculating the integral

\begin{equation}
\IA(\position^{\prime}-\position,\spin^{\prime},\spin)=\int_{-\pi}^{\pi}{\rm d}\K e^{i\K(x^{\prime}-x)}\sum_{\dspin}\langle\spin^{\prime}|\dspin(K)\rangle\langle\dspin(\K)|\spin\rangle\lambda_{\dspin}^{*}(\K)^{-2}(\partial_{\K}[\lambda_{\dspin}^{*}(\K)])^{2}.
\end{equation}
When $\spin^{\prime}=\spin$,{\small
\begin{equation}
\begin{aligned}\IA(\position^{\prime}-\position,\spin^{\prime},\spin)= & \int_{-\pi}^{\pi}[-\frac{2e^{i\K(\position^{\prime}-\position)}\cos^{2}(\K)}{\cos(2\K)+3}]{\rm d}\K\overset{z=e^{i\K}}{=}\oint_{|z|=1}[i\frac{z^{(\position^{\prime}-\position-1)}(z+z^{-1})^{2}}{((z+z^{-1})^{2}+4)}]dz=\oint_{|z|=1}i\left[z^{(\position^{\prime}-\position-1)}-\frac{4z^{(\position^{\prime}-\position+1)}}{z^{4}+6z^{2}+1}\right]\mathrm{d}z.\end{aligned}
\end{equation}
}The roots of the equation $z^{4}+6z^{2}+1=0$ are $-i(\sqrt{2}-1),i(\sqrt{2}-1),-i(\sqrt{2}+1),i(\sqrt{2}+1)$,
among which $-i(\sqrt{2}-1),i(\sqrt{2}-1)$ are inside the unit circle
on the complex plane and $-i(\sqrt{2}+1),i(\sqrt{2}+1)$ are outside.
They are the first-order poles of the integrand. In addition, $z^{(\position^{\prime}-\position-1)}$
may produce a residue when $\position^{\prime}-\position-1=-1$. Then

\begin{equation}
\begin{aligned}\IA(\position^{\prime}-\position,\spin^{\prime},\spin)= & \begin{cases}
-2\pi i\{\mathrm{Res}[f(-i\sqrt{3+2\sqrt{2}})]+\mathrm{Res}[f(i\sqrt{3+2\sqrt{2}})]\}, & \position^{\prime}-\position<0\\
+2\pi i\{\mathrm{Res}[f(-i\sqrt{3-2\sqrt{2}})]+\mathrm{Res}[f(i\sqrt{3-2\sqrt{2}})]\}-2\pi i\mathrm{Res}[f(\infty)], & \position^{\prime}-\position=0\\
+2\pi i\{\mathrm{Res}[f(-i\sqrt{3-2\sqrt{2}})]+\mathrm{Res}[f(i\sqrt{3-2\sqrt{2}})]\}, & \position^{\prime}-\position>0
\end{cases}\\
= & \begin{cases}
\sqrt{2}\pi\left(3+2\sqrt{2}\right)^{(\position^{\prime}-\position)/2}\cos\left(\frac{\pi(\position^{\prime}-\position)}{2}\right), & \position^{\prime}-\position<0\\
(\sqrt{2}-2)\pi, & \position^{\prime}-\position=0\\
\sqrt{2}\pi\left(3-2\sqrt{2}\right)^{(\position^{\prime}-\position)/2}\cos\left(\frac{\pi(\position^{\prime}-\position)}{2}\right), & \position^{\prime}-\position>0
\end{cases}\\
= & \begin{cases}
(\sqrt{2}-2)\pi, & \position^{\prime}-\position=0\\
\sqrt{2}\pi\left(3-2\sqrt{2}\right)^{|\position^{\prime}-\position|/2}\cos\left(\frac{\pi(\position^{\prime}-\position)}{2}\right), & \position^{\prime}-\position\ne0.
\end{cases}
\end{aligned}
\end{equation}
The notation $\mathrm{Res}[f(a)]$ represents the residue of function
$f(z)$ at point $a$. When $\spin^{\prime}\ne\spin$,
\begin{equation}
\IA(\position^{\prime}-\position,\spin^{\prime},\spin)=0.
\end{equation}

To sum up,
\begin{equation}
\begin{aligned}\IA(\position^{\prime}-\position,\spin^{\prime},\spin)= & \begin{cases}
(\sqrt{2}-2)\pi, & \position^{\prime}-\position=0,\spin^{\prime}=\spin\\
\sqrt{2}\pi\left(\sqrt{2}-1\right)^{|\position^{\prime}-\position|}\cos\left(\frac{\pi(\position^{\prime}-\position)}{2}\right), & \position^{\prime}-\position\ne0,\spin^{\prime}=\spin\\
0, & \spin^{\prime}\ne\spin.
\end{cases}\end{aligned}
\end{equation}

\subsection{Optimization problem of coefficient\protect\label{sec:Optimization-problem-of}}

The coefficient $c_{2}$ of $t^{2}$ in the expression of the position
variance after $t$ steps of evolution $\langle\Psi|(\evolve^{\dagger})^{t}\distanceOperator\evolve^{t}|\Psi\rangle$
can be calculated from the combination of inner products of $\langle\position_{1}\spin_{1}^{\prime}|\cdots\langle x_{n}\spin_{n}^{\prime}|(\evolven^{\dagger})^{t}\distanceOperator\evolven^{t}|\position_{1}\spin_{1}\rangle\cdots|x_{n}\spin_{n}\rangle$
{[}see Eq. \eqref{eq:coindistance}{]}. For simplicity, we regard
$\uparrow$ as number $1$ and $\downarrow$ as number $0$ and denote
the spin of the $i$th particle as $|s_{i}\rangle$. Then the vector
$|\spin_{i}\rangle$ takes on the value of $|1\rangle$ or $|0\rangle$,
and the combination of the spin notations of these particles can be
treated as a binary number whose numerical value is $\xi$, that is,
$|\spin_{1}\spin_{2}\cdots\spin_{n}\rangle=|2^{n-1}\times\spin_{1}+2^{n-2}\times\spin_{2}+\cdots+2\spin_{n-1}+\spin_{n}\rangle=|\xi\rangle$.
We can rewrite the basis state vectors as $|\position_{1}\rangle|\position_{2}\rangle\cdots|x_{n}\rangle|\xi\rangle$.

The position variance is given by
\begin{equation}
\begin{aligned}\langle\Psi|(\hat{U}^{\dagger})\hat{D}\hat{U}^{t}|\Psi\rangle= & \left[\langle\position_{1}|\langle\position_{2}|\cdots\langle x_{n}|\sum_{\xi^{\prime}=0}^{2^{n}-1}a_{\xi^{\prime}}^{*}\langle\xi^{\prime}|\right](\evolven^{\dagger})^{t}\distanceOperator\evolven^{t}\left[|\position_{1}\rangle|\position_{2}\rangle\cdots|x_{n}\rangle\sum_{\xi=0}^{2^{n}-1}a_{\xi}|\xi\rangle\right],\end{aligned}
\label{eq:coindistance}
\end{equation}
in which the coefficient of $t^{2}$ is

\begin{equation}
\begin{aligned}c_{2}= & \sum_{\xi^{\prime}\xi}a_{\xi^{\prime}}^{*}a_{\xi}M_{\xi^{\prime}\xi}=[a_{0}^{*},a_{1}^{*},...,a_{2^{n}-1}^{*}]\matrixTsquare\left[\begin{array}{c}
a_{0}\\
a_{1}\\
\vdots\\
a_{2^{n}-1}
\end{array}\right]\end{aligned}
.
\end{equation}
The matrix elements of $M$ are given by

\begin{equation}
\begin{aligned}\matrixTsquare_{\xi^{\prime}\xi}= & -\frac{n-1}{n}\frac{1}{2\pi}\sum_{i}\IA(0,\spin_{i}^{\prime},\spin_{i})\prod_{\gamma=1,\gamma\ne i}^{n}\delta_{\position_{\gamma}^{\prime}\spin_{\gamma}^{\prime},\position_{\gamma}\spin_{\gamma}}\\
 & +\frac{1}{n}\left(\frac{1}{2\pi}\right)^{2}\sum_{j,k,j\ne k}\IB(0,\spin_{j}^{\prime},\spin_{j})\IB(0,\spin_{k}^{\prime},\spin_{k})\prod_{\gamma=1,\gamma\ne j,k}^{n}\delta_{\position_{\gamma}^{\prime}\spin_{\gamma}^{\prime},\position_{\gamma}\spin_{\gamma}}.
\end{aligned}
\end{equation}
In $\IA(0,\spin_{i}^{\prime},\spin_{i})$, if $\spin_{i}^{\prime}\ne s_{i}$,
$\IA(0,\spin_{i}^{\prime},\spin_{i})=0$. Thus the first part in $M_{\xi^{\prime}\xi}$
is zero if $\xi^{\prime}\ne\xi$. If $\xi^{\prime}=\xi$, it takes
on the value of $(n-1)(1-\frac{1}{\sqrt{2}})$.

For the second part in $M_{\xi^{\prime}\xi}$, integrals $\IB(0,\uparrow,\uparrow)$,
$\IB(0,\uparrow,\downarrow)$, and $\IB(0,\downarrow,\uparrow)$ are
all $i(2-\sqrt{2})\pi$, while $\IB(0,\downarrow,\downarrow)=-i(2-\sqrt{2})\pi$.
The values of $\IB(0,\spin_{j}^{\prime},\spin_{j})\IB(0,\spin_{k}^{\prime},\spin_{k})$
corresponding to different combinations of $s_{j}^{\prime}$,$\spin_{k}^{\prime}$,$\spin_{j}$,
and $\spin_{k}$ are listed below:
\begin{center}
\begin{tabular}{|c|c|}
\hline 
$(s_{j}^{\prime},\spin_{k}^{\prime},\spin_{j},\spin_{k})$ & Value\tabularnewline
\hline 
$\begin{array}{c}
(\uparrow,\uparrow,\uparrow,\uparrow),(\uparrow,\uparrow,\uparrow,\downarrow),(\uparrow,\uparrow,\downarrow,\uparrow),(\uparrow,\uparrow,\downarrow,\downarrow),(\uparrow,\downarrow,\uparrow,\uparrow),\\
(\uparrow,\downarrow,\downarrow,\uparrow),(\downarrow,\uparrow,\uparrow,\uparrow),(\downarrow,\uparrow,\uparrow,\downarrow),(\downarrow,\downarrow,\uparrow,\uparrow),(\downarrow,\downarrow,\downarrow,\downarrow)
\end{array}$ & $-(2-\sqrt{2})^{2}\pi^{2}$\tabularnewline
\hline 
$(\uparrow,\downarrow,\uparrow,\downarrow)$,$(\uparrow,\downarrow,\downarrow,\downarrow)$,$(\downarrow,\uparrow,\downarrow,\uparrow)$,$(\downarrow,\uparrow,\downarrow,\downarrow)$,$(\downarrow,\downarrow,\uparrow,\downarrow)$,$(\downarrow,\downarrow,\downarrow,\uparrow)$ & $(2-\sqrt{2})^{2}\pi^{2}$\tabularnewline
\hline 
\end{tabular}
\par\end{center}

\begin{flushleft}
When $\xi^{\prime}=\xi$, i.e., the left state vector is the same
as the right state vector, we denote the number of particles that
are spinning up as $n_{\uparrow}$, and those spinning down as $n_{\downarrow}$.
Then $n_{\uparrow}+n_{\downarrow}=n$, and
\par\end{flushleft}

\begin{equation}
\begin{aligned}M_{\xi^{\prime}\xi}= & (n-1)\left(1-\frac{1}{\sqrt{2}}\right)+\frac{1}{n}\sum_{j\ne k}(-1)^{\delta_{s_{j},s_{k}}}\left(1-\frac{1}{\sqrt{2}}\right)^{2}\\
= & (n-1)\left(1-\frac{1}{\sqrt{2}}\right)+\frac{2}{n}[-(C_{n_{\uparrow}}^{2}+C_{n_{\downarrow}}^{2})+n_{\uparrow}n_{\downarrow}]\left(1-\frac{1}{\sqrt{2}}\right)^{2}\\
= & (n-1)\left(1-\frac{1}{\sqrt{2}}\right)-\frac{1}{n}[(n_{\uparrow}-n_{\downarrow})^{2}-n]\left(1-\frac{1}{\sqrt{2}}\right)^{2}.
\end{aligned}
\end{equation}

When $d(\xi^{\prime},\xi)>2$, $M_{\xi^{\prime}\xi}=0$, and when
$d(\xi^{\prime},\xi)=1$, there is a mismatch in one of the spins
pairs. We denote the number of spinning up pairs in the matched $n-1$
pairs as $n_{\uparrow}^{\prime}$ and those spinning down as $n_{\downarrow}^{\prime}$;
the matrix elements are
\begin{equation}
M_{\xi^{\prime}\xi}=\frac{2}{n}[-n_{\uparrow}^{\prime}+n_{\downarrow}^{\prime}]\left(1-\frac{1}{\sqrt{2}}\right)^{2}.
\end{equation}

When $d(\xi^{\prime},\xi)=2$,

\begin{equation}
M_{\xi^{\prime}\xi}=-\frac{2}{n}\left(1-\frac{1}{\sqrt{2}}\right)^{2}.
\end{equation}

If we rewrite the subscripts of matrix $M$ as $i,j,k$, then
\begin{equation}
\begin{aligned}M_{ii}= & (n-1)\left(1-\frac{1}{\sqrt{2}}\right)+\frac{1}{n}[n-(2W(i)-n)^{2}]\left(1-\frac{1}{\sqrt{2}}\right)^{2},\end{aligned}
\end{equation}
\begin{equation}
\begin{aligned}\matrixTsquare_{jk}= & \frac{2}{n}\left(1-\frac{1}{\sqrt{2}}\right)^{2}[\delta_{d(j,k),1}(n-1-2\min\{W(j),W(k)\})-\delta_{d(j,k),2}],(j\ne k),\end{aligned}
\end{equation}
where $i,j,k=0,1,2,...,2^{n}-1$ and $W(i)$ represents the Hamming
weight of integer $i$, i.e. the number of $1$'s in the $n$-bit
binary form of integer $i$. Here $d(j,k)$ represents the Hamming
distance between integers $j$ and $k$, which is the number of bit
positions where the corresponding two bits in the binary form of $i$
and $j$ are different. $\delta$ is the Kronecker delta.

The maximum and minimum values of $c_{2}$ as a function of the initial
coin state are the maximum and minimum eigenvalues of $\matrixTsquare$,
respectively.

\subsection{Solving eigenvalues and eigenvectors\protect\label{sec:Solving-eigenvalues-and}}

Given the number of particles, $n$, the matrix elements of $M$ are

\begin{equation}
\begin{aligned} & M_{ii}=(n-1)\left(1-\frac{1}{\sqrt{2}}\right)+\frac{1}{n}[n-(2W(i)-n)^{2}]\left(1-\frac{1}{\sqrt{2}}\right)^{2}\\
 & M_{jk}=\frac{2}{n}\left(1-\frac{1}{\sqrt{2}}\right)^{2}[\delta_{d(j,k),1}(n-1-2\min\{W(j),W(k)\})-\delta_{d(j,k),2}],(j\ne k),
\end{aligned}
\end{equation}
where $i,j,k\in\{0,1,...,2^{n}-1\}$. We define

\begin{equation}
\B_{n}=\left(1-\frac{1}{\sqrt{2}}\right)^{-2}\frac{n}{2}\matrixTsquare_{n}-\frac{n(n-1)}{\sqrt{2}}I,
\end{equation}
the matrix elements of which are

\begin{equation}
\begin{aligned} & b_{ii}=n(n-1)+\frac{1}{2}[n-(2W(i)-n)^{2}]\\
 & b_{jk}=\delta_{d(j,k),1}(n-1-2\min\{W(j),W(k)\})-\delta_{d(j,k),2},(j\ne k).
\end{aligned}
\end{equation}

When $n\ge2$, we define the $2\times2$ matrix $P_{ij}=(-1)^{c(i,j)}\Dmat^{d(i,j)}$,
where
\begin{equation}
\Dmat=\left[\begin{array}{cc}
0 & 1\\
1 & 2
\end{array}\right],
\end{equation}
$i,j\in\{0,1,...,2^{n}-1\}$ are even; $c(i,j)$ represents the number
of bit positions where the corresponding two bits in the binary form
of $i$ and $j$ are all ones.

\newtheorem{theorem}{Theorem}
\begin{theorem}

\label{theorem:eigen}Denote $i,j$ to be the row and column locations
of the upper-left corner element of the submatrix $P_{ij}$ in the
$2^{n}\times2^{n}$ matrix $P$. The submatrices $P_{ij}$ make up
the matrix $P_{n}$, such that $P_{n}^{-1}\B_{n}P_{n}$ is diagonal
matrix $(\mu_{k}I_{2}),$ where $k$ is an even number and $I_{2}$
is a $2\times2$ identity matrix.

\end{theorem}

\begin{proof}

When $n=2$,

\begin{equation}
P_{n}^{-1}B_{n}P_{n}=\left[\begin{array}{cccc}
0 & 0 & 0 & 0\\
0 & 0 & 0 & 0\\
0 & 0 & 4 & 0\\
0 & 0 & 0 & 4
\end{array}\right].
\end{equation}

We prove the theorem by mathematical induction. Assume that it holds
true for $n$. Then for $n+1$, let $B_{n}=(B_{ij})$ be the $2\times2$
submatrix partition and similarly $B_{n+1}=(B_{ij}^{\prime}),$ $P_{n}=(P_{ij})$,
$P_{n+1}=(P_{ij}^{\prime})$, where $i,j$ are even numbers. It can
be proved that $\forall i,k<2^{n+1}$, $2|i,2|k$, $\exists\mu^{\prime}$
such that 

\begin{equation}
\sum_{\substack{0\le j<2^{n+1}\\
2|j
}
}B_{ij}^{\prime}P_{jk}^{\prime}=\mu^{\prime}P_{ik}^{\prime}.
\end{equation}

We first prove it for the case of $i<2^{n},k<2^{n}$. Let $j<2^{n}$,
\begin{equation}
B_{ij}^{\prime}=B_{ij}+\left[\begin{array}{cc}
\delta_{d(i,j),1} & \delta_{d(i,j+1),1}\\
\delta_{d(i,j+1),1} & \delta_{d(i+1,j+1),1}
\end{array}\right]=B_{ij}+\delta_{d(i,j),1}I_{2}(i\ne j),
\end{equation}
\begin{equation}
B_{ii}^{\prime}=B_{ii}+\left[\begin{array}{cc}
n+2W(i) & 1\\
1 & n+2+2W(i)
\end{array}\right],
\end{equation}
\begin{equation}
B_{i,j+2^{n}}^{\prime}=\left[\begin{array}{cc}
-\delta_{d(i,j+2^{n}),2} & -\delta_{d(i,j+2^{n}+1),2}\\
-\delta_{d(i+1,j+2^{n}),2} & -\delta_{d(i+1,j+2^{n}+1),2}
\end{array}\right]=-\delta_{d(i,j),1}I_{2}(i\ne j),
\end{equation}
\begin{equation}
B_{i,i+2^{n}}^{\prime}=\left[\begin{array}{cc}
n-2W(i) & -1\\
-1 & n-2-2W(i)
\end{array}\right].
\end{equation}

Hence,
\begin{equation}
\begin{aligned}\sum_{\substack{0\le j<2^{n+1}\\
2|j
}
}B_{ij}^{\prime}P_{jk}^{\prime}= & \sum_{\substack{j<2^{n}\\
2|j\ne i
}
}(B_{ij}+\delta_{d(i,j),1}I)P_{jk}+\left[B_{ii}+\left(\begin{array}{cc}
n+2W(i) & 1\\
1 & n+2+2W(i)
\end{array}\right)\right]P_{ik}\\
+ & \sum_{\substack{j<2^{n}\\
2|j\ne i
}
}[-\delta_{d(i,j),1}\Dmat P_{jk}]+\left(\begin{array}{cc}
n-2W(i) & -1\\
-1 & n-2-2W(i)
\end{array}\right)\Dmat P_{ik}.
\end{aligned}
\end{equation}

Utilizing the induction hypothesis, i.e. $\exists\mu$ such that 
$\sum_{\substack{j<2^{n}\\
2|j
}
}B_{ij}P_{jk}=\mu P_{ik}$, we obtain

{\small
\begin{equation}
\begin{aligned}\sum_{\substack{0\le j<2^{n+1}\\
2|j
}
}B_{ij}^{\prime}P_{jk}^{\prime}= & \sum_{\substack{j<2^{n}\\
2|j\ne i
}
}\delta_{d(i,j),1}P_{jk}(I-\Dmat)+\left[\left(\begin{array}{cc}
n+2W(i) & 1\\
1 & n+2+2W(i)
\end{array}\right)+\left(\begin{array}{cc}
n-2W(i) & -1\\
-1 & n-2-2W(i)
\end{array}\right)\Dmat\right]P_{ik}+\mu P_{ik}.\end{aligned}
\label{eq:A64}
\end{equation}
}{\small\par}

Here we consider $\sum_{\substack{j<2^{n}\\
2|j,d(i,j)=1
}
}P_{jk}P_{ik}^{-1}=rI+y\Dmat$, where $r=2\times\#\{j|(i,k)_{\vartheta(i,j)}=(0,1)\}-2\times\#\{j|(i,k)_{\vartheta(i,j)}=(1,0)\}$,
$y=\#\{j|(i,k)_{\vartheta(i,j)}=(0\ \mathrm{or}\ 1,0)\}-\#\{j|(i,k)_{\vartheta(i,j)}=(0\ \mathrm{or}\ 1,1)\}$.
$\#\{j|(i,k)_{\vartheta(i,j)}=(0,1)\}$ represents the number of $j$'s
for $0\le j<2^{n},2|j$, $d(i,j)=1$ such that  the binary form of
$i$ is 0 and $k$ is 1 on the bit where the binary form of $i$ differs
from $j$. The notation $\vartheta(i,j)$ denotes the bit position
where $i$ is different from $j$. One can deduce that $r=2(W(k)-W(i))$,
$y=n-2W(k)-1$. Taking $\mu^{\prime}=\mu+4W(k)$, it can be verified
that Eq. \eqref{eq:A64} can be turned into{\small
\begin{equation}
\begin{aligned}\sum_{\substack{0\le j<2^{n+1}\\
2|j
}
}B_{ij}^{\prime}P_{jk}^{\prime}= & (rI+y\Dmat)(I-\Dmat)P_{ik}+\left[\left(\begin{array}{cc}
n+2W(i) & 1\\
1 & n+2+2W(i)
\end{array}\right)+\left(\begin{array}{cc}
n-2W(i) & -1\\
-1 & n-2-2W(i)
\end{array}\right)\Dmat\right]P_{ik}+\mu P_{ik}\\
= & [(rI+y\Dmat)(I-\Dmat)+(n+2W(i)-1)I+(n-2W(i)-1)\Dmat+\mu I]P_{ik}=\mu^{\prime}P_{ik}=\mu^{\prime}P_{ik}^{\prime}.
\end{aligned}
\end{equation}
}Note that $P_{ik}$ commutes with $\Dmat$.

When $i<2^{n}$, $k\ge2^{n}$, we denote $k^{\prime}=k-2^{n}$, so
$0\le k^{\prime}<2^{n}$. Similarly,
\begin{equation}
\begin{aligned}\sum_{\substack{0\le j<2^{n+1}\\
2|j
}
}B_{ij}^{\prime}P_{jk}^{\prime}= & \sum_{\substack{j<2^{n}\\
2|j\ne i
}
}(B_{ij}+\delta_{d(i,j),1}I)\Dmat P_{jk^{\prime}}+\left[B_{ii}+\left(\begin{array}{cc}
n+2W(i) & 1\\
1 & n+2+2W(i)
\end{array}\right)\right]\Dmat P_{ik^{\prime}}\\
+ & \sum_{\substack{j<2^{n}\\
2|j\ne i
}
}\delta_{d(i,j),1}P_{jk^{\prime}}-\left(\begin{array}{cc}
n-2W(i) & -1\\
-1 & n-2-2W(i)
\end{array}\right)P_{ik^{\prime}}.
\end{aligned}
\end{equation}
Hence, taking $\mu^{\prime}=4n-4W(k^{\prime})+\mu$, we obtain {\small
\begin{equation}
\begin{aligned}\sum_{\substack{0\le j<2^{n+1}\\
2|j
}
}B_{ij}^{\prime}P_{jk}^{\prime}= & (rI+y\Dmat)(\Dmat+I)P_{ik^{\prime}}+\left[\left(\begin{array}{cc}
n+2W(i) & 1\\
1 & n+2+2W(i)
\end{array}\right)\Dmat-\left(\begin{array}{cc}
n-2W(i) & -1\\
-1 & n-2-2W(i)
\end{array}\right)\right]P_{ik^{\prime}}+\mu\Dmat P_{ik^{\prime}}\\
= & [4(n-W(k^{\prime}))+\mu]P_{ik}^{\prime}=\mu^{\prime}P_{ik}^{\prime}.
\end{aligned}
\end{equation}
}{\small\par}

When $i\ge2^{n}$, $k<2^{n}$, we let $j<2^{n}$ and $i^{\prime}=i-2^{n}$,
\begin{equation}
B_{i^{\prime}+2^{n},j}^{\prime}=B_{j,i^{\prime}+2^{n}}^{\prime}=\left(\begin{array}{cc}
-\delta_{d(j,i^{\prime}+2^{n}),2} & -\delta_{d(j,i^{\prime}+2^{n}+1),2}\\
-\delta_{d(j+1,i^{\prime}+2^{n}),2} & -\delta_{d(j+1,i^{\prime}+2^{n}+1),2}
\end{array}\right)=-\delta_{d(i^{\prime},j),1}I_{2}(i^{\prime}\ne j),
\end{equation}
\begin{equation}
B_{i^{\prime}+2^{n},i^{\prime}}^{\prime}=\left(\begin{array}{cc}
n-2W(i^{\prime}) & -1\\
-1 & n-2-2W(i^{\prime})
\end{array}\right),
\end{equation}
\begin{equation}
B_{i^{\prime}+2^{n},j+2^{n}}^{\prime}=\left(\begin{array}{cc}
-2\delta_{d(i^{\prime},j),1} & -2\delta_{d(i^{\prime},j+1),1}\\
-2\delta_{d(i^{\prime}+1,j),1} & -2\delta_{d(i^{\prime}+1,j+1),1}
\end{array}\right)+B_{i^{\prime}j}^{\prime}=B_{i^{\prime}j}-\delta_{d(i^{\prime},j),1}I_{2}(i^{\prime}\ne j),
\end{equation}
\begin{equation}
B_{i^{\prime}+2^{n},i^{\prime}+2^{n}}^{\prime}=\left(\begin{array}{cc}
-4W(i^{\prime})+2n & -2\\
-2 & -4-4W(i^{\prime})+2n
\end{array}\right)+B_{i^{\prime}i^{\prime}}^{\prime}=B_{i^{\prime}i^{\prime}}+\left(\begin{array}{cc}
3n-2W(i^{\prime}) & -1\\
-1 & 3n-2W(i^{\prime})-2
\end{array}\right).
\end{equation}
It can also be deduced that 
\begin{equation}
\begin{aligned}\sum_{\substack{0\le j<2^{n+1}\\
2|j
}
}B_{i^{\prime}+2^{n},j}^{\prime}P_{jk}^{\prime}= & \sum_{\substack{j<2^{n}\\
2|j\ne i^{\prime}
}
}(-\delta_{d(i^{\prime},j),1})P_{jk}+\left(\begin{array}{cc}
n-2W(i^{\prime}) & -1\\
-1 & n-2-2W(i^{\prime})
\end{array}\right)P_{i^{\prime}k}\\
+ & \sum_{\substack{j<2^{n}\\
2|j\ne i^{\prime}
}
}(B_{i^{\prime}j}-\delta_{d(i^{\prime},j),1})\Dmat P_{jk}+\left[B_{i^{\prime}i^{\prime}}+\left(\begin{array}{cc}
3n-2W(i^{\prime}) & -1\\
-1 & 3n-2W(i^{\prime})-2
\end{array}\right)\right]\Dmat P_{i^{\prime}k}\\
= & [4W(k)+\mu]P_{i,k}^{\prime}.
\end{aligned}
\end{equation}

When $i\ge2^{n}$, $k\ge2^{n}$, we let $k^{\prime}=k-2^{n}$, $i^{\prime}=i-2^{n}$. Then
\begin{equation}
\begin{aligned}\sum_{\substack{0\le j<2^{n+1}\\
2|j
}
}B_{i^{\prime}+2^{n},j}^{\prime}P_{j,k^{\prime}+2^{n}}^{\prime}= & \sum_{\substack{j<2^{n}\\
2|j\ne i^{\prime}
}
}(-\delta_{d(i^{\prime},j),1})\Dmat P_{jk^{\prime}}+\left(\begin{array}{cc}
n-2W(i^{\prime}) & -1\\
-1 & n-2-2W(i^{\prime})
\end{array}\right)\Dmat P_{i^{\prime}k^{\prime}}\\
- & \sum_{\substack{j<2^{n}\\
2|j\ne i^{\prime}
}
}(B_{i^{\prime}j}-\delta_{d(i^{\prime},j),1})P_{jk^{\prime}}-[B_{i^{\prime}i^{\prime}}+\left(\begin{array}{cc}
3n-2W(i^{\prime}) & -1\\
-1 & 3n-2W(i^{\prime})-2
\end{array}\right)]P_{i^{\prime}k^{\prime}}\\
= & [4n-4W(k^{\prime})+\mu]P_{i,k}^{\prime}.
\end{aligned}
\end{equation}
\qedhere 

\end{proof}

According to Theorem \ref{theorem:eigen}, the eigenvalue corresponding
to the column subscript $k$ is 
\begin{equation}
\mu_{k}^{n}=\begin{cases}
\mu_{k}^{n-1}+4W(k), & 0\le k<2^{n-1}\\
4(n-1)-4W(k-2^{n-1})+\mu_{k-2^{n-1}}^{n-1}, & 2^{n-1}\le k<2^{n},
\end{cases}
\end{equation}
where $k$ is an even number. As $\mu_{0}^{2}=0$ and $\mu_{2}^{2}=4$,
for $n$ particles, we have
\begin{equation}
\mu_{k}=4W(k)W(2^{n}-1-k).
\end{equation}
Each $\mu_{k}$ ($0\le k<2^{n}$, $2|k$) corresponds to two eigenvectors,
and the corresponding eigenvalue of $\matrixTsquare_{n}$ is $\eta_{k}=\left(1-\frac{1}{\sqrt{2}}\right)^{2}[\frac{2\mu_{k}}{n}+(n-1)\sqrt{2}]$.
By the range of $k$, it can be inferred that $0\le W(k)\le n-1$,
$W(k)\in\mathbb{Z}$, and there are $C_{n-1}^{W(k)}$ different $k$'s
that correspond to the same value of $W(k)$. Note that for different
$k$'s, e.g., $k$ and $k^{\prime}$, $\eta_{k}$ and $\eta_{k^{\prime}}$
are the same when $W(k)=n-W(k^{\prime})$, which should be taken into
account in counting the degeneracies of the eigenvalues. We assume
that $k,k^{\prime}\in\{0,2,4,...,2^{n}-2\}$. When $W(k)=0$, there
is no $k^{\prime}$ satisfying $W(k^{\prime})=n-W(k)=n$. When $n$
is odd, for any given $k$ satisfying $1\le W(k)\le(n-1)/2$, there
are $k^{\prime}$'s $(k^{\prime}\ne k)$ such that $(n+1)/2\le W(k^{\prime})\le n-1$
and $W(k^{\prime})=n-W(k)$. When $n$ is even, for any given $k$
satisfying $1\le W(k)\le n/2-1$, there is another set of $k^{\prime}$'s
$(k^{\prime}\ne k)$ such that $n/2+1\le W(k^{\prime})\le n-1$ and
$W(k^{\prime})=n-W(k)$, and for those $k$'s satisfying $W(k)=n/2$,
as $n-W(k)=W(k)$, the degeneracy of $\eta_{k}$ is $2C_{n-1}^{W(k)}$.
So we arrive at the following result for the degeneracy of $\eta_{k}$:
\begin{equation}
\mathrm{count}(\eta_{k})=\begin{cases}
2, & k=0\\
2(C_{n-1}^{W(k)}+C_{n-1}^{n-W(k)}), & k\ne0,n\mathrm{\ is\ odd}\\
2(C_{n-1}^{W(k)}+C_{n-1}^{n-W(k)}), & k\ne0,W(k)\ne\frac{n}{2},n\mathrm{\ is\ even}\\
2C_{n-1}^{W(k)}, & W(k)=\frac{n}{2},n\mathrm{\ is\ even}.
\end{cases}
\end{equation}

\subsection{Proof of partial exchange symmetry\protect\label{sec:Proof-of-partial}}

\global\long\def\rowindex{i^{\prime}}%
\global\long\def\columnindex{j^{\prime}}%

Using the relation $\Dmat^{2}=2\Dmat+I$, it can be proven that
\begin{equation}
\Dmat^{\alpha}=\left[\begin{array}{cc}
\F(\alpha-1) & \F(\alpha)\\
\F(\alpha) & \F(\alpha+1)
\end{array}\right],
\end{equation}
where $\alpha\in\mathbb{Z}$ and
\begin{equation}
\F(\alpha)=\frac{(1+\sqrt{2})^{\alpha}-(1-\sqrt{2})^{\alpha}}{2\sqrt{2}}.
\end{equation}
$F(\alpha)\ge0$ when $\alpha\ge-1$. So,
\begin{equation}
P_{ij}=(-1)^{c(i,j)}\Dmat^{d(i,j)}=(-1)^{c(i,j)}\left[\begin{array}{cc}
F(d(i,j)-1) & F(d(i,j))\\
F(d(i,j)) & F(d(i,j)+1)
\end{array}\right].
\end{equation}

To analyze the exchange symmetry, we take a close look at the binary
form of subscripts $i$ and $j$, that $i,j\in\{0,2,...,2^{n}-2\}$.
Each column of the matrix $P_{n}=(P_{ij})$ is an unnormalized eigenvector
of matrix $M$. Combining the normalized eigenvectors and the position
state, we obtain the optimized initial states for the quantum walk.
Let $\rowindex,\columnindex\in\{0,1,...,2^{n}-1\}$ be the row and
column subscripts of the matrix element $p_{i^{\prime}j^{\prime}}$
in $P_{n}$. Then the element in row $\rowindex=(\rowindex_{1}\rowindex_{2}\cdots\rowindex_{n})_{2}$
of a normalized eigenvector is the coefficient with respect to the
product basis $|\rowindex\rangle=|\rowindex_{1}\rangle|\rowindex_{2}\rangle\cdots|\rowindex_{n}\rangle$,
where $|i_{\beta}^{\prime}\rangle$ corresponds to the $\beta$th
particle. Let $P_{ij}$ be the $2\times2$ submatrix of $P_{n}$ that
the matrix element $p_{i^{\prime}j^{\prime}}$ belongs to, and the
row and column locations of the upper-left element of $P_{ij}$ are
$i=2\lfloor i^{\prime}/2\rfloor$, $j=2\lfloor j^{\prime}/2\rfloor$.
We will see that each eigenstate is invariant under the permutation
of some of the $n$ particles, and we define this invariance of the
eigenstates as \emph{partial exchange symmetry}. When the $u$th and
$u^{\prime}$th particles are swapped, the state vector is changed
by exchanging the single-particle states of the $u$th and $u^{\prime}$th
particles in all the product basis states that compose the whole multiparticle
state. 

Consider an initial state with the coin state corresponding to the
$j^{\prime}$th column of matrix $P_{n}$ as
\begin{equation}
|\Psi^{(j^{\prime})}\rangle=|00\cdots0\rangle\otimes\sum_{i^{\prime}}a_{i^{\prime}}^{(j^{\prime})}|i^{\prime}\rangle,
\end{equation}
where $|i^{\prime}\rangle=|i_{1}^{\prime}i_{2}^{\prime}\cdots i_{n}^{\prime}\rangle$,
and $a_{i^{\prime}}^{(j^{\prime})}=p_{i^{\prime}j^{\prime}}/\sqrt{\sum_{i^{\prime\prime}}\left|p_{i^{\prime\prime}j^{\prime}}\right|^{2}}$.
We define a map $g_{\beta,\gamma}:\{0,1,...,2^{n}-1\}\rightarrow\{0,1,...,2^{n}-1\}$
that maps a number to another whose numbers on bit positions $\beta$
and $\gamma$ of the binary representation are swapped, e.g., $g_{\beta,\gamma}((i_{1}\cdots i_{\beta}\cdots i_{\gamma}\cdots i_{n})_{2})=(i_{1}\cdots i_{\gamma}\cdots i_{\beta}\cdots i_{n})_{2}$.
Since the initial position states of the particles are all $|0\rangle$,
the symmetry of the state $|\Psi^{(j^{\prime})}\rangle$ is completely
determined by its multiparticle coin state. Then the invariance of
the state with respect to exchanging the $\beta$th and $\gamma$th
particles is equivalent to $\sum_{i^{\prime}}a_{i^{\prime}}^{(j^{\prime})}|i^{\prime}\rangle=\sum_{i^{\prime}}a_{i^{\prime}}^{(j^{\prime})}|g_{\beta,\gamma}(i^{\prime})\rangle$,
i.e. $a_{i^{\prime}}^{(j^{\prime})}=a_{g_{\beta,\gamma}(i^{\prime})}^{(j^{\prime})}$
and thus $p_{i^{\prime}j^{\prime}}=p_{g_{\beta,\gamma}(i^{\prime}),j^{\prime}}$
for all $i^{\prime}$. We define another map $h_{\beta,\gamma}:\{0,1,...,2^{n}-1\}\rightarrow\{0,2,...,2^{n}-2\}$,
such that $h_{\beta,\gamma}(i^{\prime})=2\lfloor g_{\beta,\gamma}(i^{\prime})/2\rfloor$.
Then $h_{\beta,\gamma}(i^{\prime})$ is the row subscript of the upper-left
corner of the submatrix that the matrix element $p_{i^{\prime}j^{\prime}}$
is mapped to by exchanging particles $\beta$ and $\gamma$. We will
omit the subscripts $\beta$ and $\gamma$ in $g_{\beta,\gamma}$
and $h_{\beta,\gamma}$ for simplicity in the following proof.

\newtheorem{lemma}{Lemma}
\begin{lemma}

\label{lemma:lemma1}If a swap in the binary form of $\rowindex$
occurs between two of the first $n-1$ bits where the corresponding
bits in $j$ are both $0$ or $1$, then $c(i,j)$ and $d(i,j)$ are
unchanged for all $i^{\prime}$, i.e. $c(h(i^{\prime}),j)=c(i,j)$,
$d(h(i^{\prime}),j)=d(i,j)$.

\end{lemma}

\begin{lemma}\label{lemma:lemma2}

If a swap in the binary form of $\rowindex$ occurs between two of
the first $n-1$ bits where the corresponding bits in $j$ are $1$
and $0$ respectively, there always exists an $\rowindex$ such that
 both $c(i,j)$ and $d(i,j)$ are changed, i.e. $c(h(i^{\prime}),j)\ne c(i,j)$
and $d(h(i^{\prime}),j)\ne d(i,j)$.

\end{lemma}

\begin{lemma}\label{lemma:lemma3}

When a swap occurs between two of the first $n-1$ particles, the
state is unchanged if and only if the particles corresponding to the
bit positions where the bits in $j$ are both $0$ or 1 are swapped,
as in the way described in Lemma \ref{lemma:lemma1}.

\end{lemma}

\begin{proof}

The sufficiency is evident. The following is the proof of necessity.
If one does not perform the swap as described in Lemma \ref{lemma:lemma1},
 but performs a swap as described in Lemma \ref{lemma:lemma2} that
the two particles to be swapped correspond to the bit positions where
the bits in $j$ are $0$ and 1 respectively, there exists an $\rowindex$,
such that  $c(h(i^{\prime}),j)=c(i,j)\pm1$ and $d(h(i^{\prime}),j)=d(i,j)\pm2$,
so the element $p_{g(i^{\prime}),j^{\prime}}$ in the submatrix $P_{h(i^{\prime}),j}=(-1)^{c(h(i^{\prime}),j)}Z^{d(h(i^{\prime}),j)}$
that $p_{i^{\prime}j^{\prime}}$ is mapped to would be different from
$p_{i^{\prime}j^{\prime}}$. Specifically, if $i^{\prime}$ is equal
to $j$ on the two bit positions to be swapped, respectively, then
$c(h(i^{\prime}),j)=c(i,j)-1$ and $d(h(i^{\prime}),j)=d(i,j)+2$,
and if $i^{\prime}$ is different from $j$ on both of the two bit
positions to be swapped, then $c(h(i^{\prime}),j)=c(i,j)+1$ and $d(h(i^{\prime}),j)=d(i,j)-2$.
As $d(h(i^{\prime}),j)\ge0$, all of the elements in matrix $Z^{d(h(i^{\prime}),j)}$
are greater than or equal to zero. So $p_{i^{\prime}j^{\prime}}$
and $p_{g(i^{\prime}),j^{\prime}}$ would be at least different in
sign due to the change of $c(i,j)$ into $c(h(i^{\prime}),j)$ in
these situations, except for those $i^{\prime}$'s such that  $p_{i^{\prime}j^{\prime}}=0$.
But it can be proven that $p_{g(i^{\prime}),j^{\prime}}$ must be
nonzero if $p_{i^{\prime}j^{\prime}}=0$. To conclude, if $j_{\beta}\ne j_{\gamma}$,
$1\le\beta,\gamma\le n-1$, $\exists i^{\prime}$ such that  $p_{i^{\prime}j^{\prime}}\ne p_{g_{\beta,\gamma}(i^{\prime}),j^{\prime}}$,
so the state is changed. \qedhere

\end{proof}

\begin{lemma}\label{lemma:lemma4}

When swapping the particle corresponding to the last bit with the
one corresponding to one of the first $n-1$ bits, the state is unchanged
if and only if the particle corresponding to the last bit is swapped
with the particle corresponding to the bit position where the corresponding
bit in $j$ is $0$.

\end{lemma}

\begin{proof}

\emph{Proof of sufficiency:} If the particle corresponding to the
last bit is swapped with the particle corresponding to one of the
first $n-1$ bit positions where the corresponding bit in $j$ is
$0$, i.e., we swap the $\beta$th and $\gamma$th particles such
that  $\beta=n$, $1\le\gamma\le n-1$, $j_{\gamma}=0$, we will show
that the coefficients of all of the bases remain unchanged after the
swap, i.e. $a_{i^{\prime}}^{(j^{\prime})}=a_{g_{\beta,\gamma}(i^{\prime})}^{(j^{\prime})}$.

As $j_{\gamma}=0$, $c(h(i^{\prime}),j)=c(i,j)$ holds true for all
$i^{\prime}\in\{0,1,...,2^{n}-1\}$. For each $i^{\prime}$ with $i_{\beta}^{\prime}=\rowindex_{n}=0$,
if $i_{\gamma}^{\prime}$ is $0$, the swap does not change $\rowindex$;
i.e., the corresponding basis is unchanged. But when $i_{\gamma}^{\prime}$
is $1$, $d(h(i^{\prime}),j)=d(i,j)-1$, and the corresponding submatrix
where $p_{g(i^{\prime}),j^{\prime}}$ locates is
\begin{equation}
P_{h(i^{\prime}),j}=(-1)^{c(i,j)}\Dmat^{d(i,j)-1}=(-1)^{c(i,j)}\left[\begin{array}{cc}
F(d(i,j)-2) & F(d(i,j)-1)\\
F(d(i,j)-1) & F(d(i,j))
\end{array}\right],
\end{equation}
 while the original submatrix $P_{ij}$ where the matrix element $p_{i^{\prime}j^{\prime}}$
locates is given by
\begin{equation}
P_{ij}=(-1)^{c(i,j)}\left[\begin{array}{cc}
F(d(i,j)-1) & F(d(i,j))\\
F(d(i,j)) & F(d(i,j)+1)
\end{array}\right].
\end{equation}
We can see that the first row of $P_{ij}$ where $p_{i^{\prime}j^{\prime}}$
locates is identical to the second row of $P_{h(i^{\prime}),j}$ where
$p_{g(i^{\prime}),j^{\prime}}$ locates, so if we modify a basis where
the last bit is $\rowindex_{n}=0$, it is mapped to another basis
in which the coefficient is the same as that in the original basis,
i.e. $p_{i^{\prime},j^{\prime}}=p_{g(i^{\prime}),j^{\prime}}$ for
all $i^{\prime}$ with $i_{n}^{\prime}=0$.

For each $i^{\prime}$ with $\rowindex_{n}=1$, the basis is unchanged
if the bit $i_{\gamma}^{\prime}$ which is swapped with $\rowindex_{n}$
is $1$. If the bit $i_{\gamma}^{\prime}$ is 0, then $d(h(i^{\prime}),j)=d(i,j)+1$,
\begin{equation}
P_{h(i^{\prime}),j}=(-1)^{c(i,j)}\left[\begin{array}{cc}
F(d(i,j)) & F(d(i,j)+1)\\
F(d(i,j)+1) & F(d(i,j)+2)
\end{array}\right].
\end{equation}
The second row of $P_{ij}$ is the same as the first row of $P_{h(i^{\prime}),j}$,
hence $p_{i^{\prime}j^{\prime}}=p_{g(i^{\prime}),j^{\prime}}$.

\emph{Proof of necessity:} If the particle corresponding to the last
bit is swapped with the particle corresponding to one of the first
$n-1$ bit positions where the corresponding bit in $j$ is $1$,
i.e., swapping the $\beta$th and $\gamma$th particles such that
$\beta=n$, $1\le\gamma\le n-1$, $j_{\gamma}=1$, there always exists
$i^{\prime}$ such that $c(h(i^{\prime}),j)=c(i,j)-1$ and $d(h(i^{\prime}),j)=d(i,j)+1$
when $\rowindex_{n}=0$ and $i_{\gamma}^{\prime}=1$, or reversely
$c(h(i^{\prime}),j)=c(i,j)+1$ and $d(h(i^{\prime}),j)=d(i,j)-1$
when $\rowindex_{n}=1$ and $i_{\gamma}^{\prime}=0$. So the corresponding
$p_{i^{\prime}j^{\prime}}$ and $p_{g(i^{\prime}),j^{\prime}}$ are
at least different in sign due to the change of $c(i,j)$ into $c(h(i^{\prime}),j)$
after the swap, except for those $i^{\prime}$'s such that  $p_{i^{\prime}j^{\prime}}=0$,
similar to the situation discussed in the proof of Lemma \ref{lemma:lemma3}.
By comparing the first row of $P_{ij}$ with the second row of $P_{h(i^{\prime}),j}$
when $i_{n}^{\prime}=0,$ $i_{\gamma}^{\prime}=1$ and comparing the
second row of $P_{ij}$ with the first row of $P_{h(i^{\prime}),j}$
when $i_{n}^{\prime}=1,i_{\gamma}^{\prime}=0$, we can also prove
that $p_{i^{\prime}j^{\prime}}$ and $p_{g(i^{\prime}),j^{\prime}}$
cannot be simultaneously $0$. Thus the state is changed if the particle
corresponding to the last bit position is swapped with the particle
corresponding to the bit position where the corresponding bit in $j$
is $1$.\qedhere

\end{proof}

Based on Lemmas \ref{lemma:lemma1}-\ref{lemma:lemma4}, we are led
to the following theorem:

\begin{theorem}

Given an initial state $|\Psi^{(j^{\prime})}\rangle$, the state
is unchanged if and only if the swap happens among some of the first
$n-1$ particles where the corresponding bits in $j=2\lfloor j^{\prime}/2\rfloor$
are all $1$ or $0$, or between the last particle and one of the
first $n-1$ particles where the corresponding bit in $j$ is 0.

\end{theorem}

So we can group the particles whose corresponding bits in $j$ are
$1$ into a set and the other particles into another. Then we connect
each pair of particles with an undirected edge if the state is unchanged
after they are swapped. Each of the two subsets (when $j\ne0$) of
particles makes up a complete graph. If we denote $n_{\uparrow}$
and $n_{\downarrow}$ as the number of $1$'s and $0$'s in the binary
form of $j$ respectively ($j$ is even), the number of possible ways
to exchange two particles preserving an eigenstate is $p_{j}=\frac{n_{\uparrow}(n_{\uparrow}-1)}{2}+\frac{n_{\downarrow}(n_{\downarrow}-1)}{2}$,
$n_{\uparrow}+n_{\downarrow}=n$, and the two disconnected complete
subgraphs have $n_{\uparrow}$ and $n_{\downarrow}$ vertices, respectively.
We can rewrite this using the Hamming weight, $p_{j}=\frac{W(j)(W(j)-1)}{2}+\frac{(n-W(j))(n-W(j)-1)}{2}=W(j)^{2}-nW(j)+C_{n}^{2}$,
so $\eta_{j}=\left(1-1/\sqrt{2}\right)^{2}\left[8(C_{n}^{2}-p_{j})/n+(n-1)\sqrt{2}\right]$.
When $j=0$, all the bits in $j$ are $0$, so there is only one complete
graph and all of the particles are vertices of this graph. As the
evolution operator $\hat{U}_{n}$ is symmetric with respect to particles,
the symmetry of the state is preserved during the evolution.

\subsection{Entanglement between complete graphs\protect\label{sec:Entanglement-between-complete}}

Since the particles in each of the two complete graphs possess exchange
symmetry, for simplicity, the eigenstates can be represented by the
superpositions of the direct products of bosonic Fock states. When
$k$ is an even number, the $k$th eigenstate is
\begin{equation}
\begin{aligned}|\eta_{k}\rangle= & C\sum_{\substack{0\le w_{1}\le W(k)\\
0\le w_{0}\le n-W(k)
}
}(-1)^{w_{1}}F(W(k)-w_{1}+w_{0}-1)\sqrt{\frac{(n-W(k))!}{w_{0}!(n-W(k)-w_{0})!}}\sqrt{\frac{W(k)!}{w_{1}!(W(k)-w_{1})!}}\\
 & \times|1_{w_{0}}0_{n-W(k)-w_{0}}\rangle\otimes|1_{w_{1}}0_{W(k)-w_{1}}\rangle\\
= & C\bigg[\frac{(1+\sqrt{2})^{W(k)-1}}{2\sqrt{2}}\sum_{0\le w_{0}\le n-W(k)}(1+\sqrt{2})^{w_{0}}\sqrt{\frac{(n-W(k))!}{w_{0}!(n-W(k)-w_{0})!}}|1_{w_{0}}0_{n-W(k)-w_{0}}\rangle\\
 & \otimes\sum_{0\le w_{1}\le W(k)}(-1)^{w_{1}}(1+\sqrt{2})^{-w_{1}}\sqrt{\frac{W(k)!}{w_{1}!(W(k)-w_{1})!}}|1_{w_{1}}0_{W(k)-w_{1}}\rangle\\
 & -\frac{(1-\sqrt{2})^{W(k)-1}}{2\sqrt{2}}\sum_{0\le w_{0}\le n-W(k)}(1-\sqrt{2})^{w_{0}}\sqrt{\frac{(n-W(k))!}{w_{0}!(n-W(k)-w_{0})!}}|1_{w_{0}}0_{n-W(k)-w_{0}}\rangle\\
 & \otimes\sum_{0\le w_{1}\le W(k)}(-1)^{w_{1}}(1-\sqrt{2})^{-w_{1}}\sqrt{\frac{W(k)!}{w_{1}!(W(k)-w_{1})!}}|1_{w_{1}}0_{W(k)-w_{1}}\rangle\bigg].
\end{aligned}
\end{equation}
where \{$|1_{w_{0}}0_{n-W(k)-w_{0}}\rangle|0\le w_{0}\le n-W(k)\}$
and $\{|1_{w_{1}}0_{W(k)-w_{1}}\rangle|0\le w_{1}\le W(k)\}$ are
two sets of orthonormal vectors. $C$ is the normalization factor.

We denote 
\begin{equation}
|\tilde{\varphi}_{A}^{1}\rangle\equiv\sum_{0\le w_{0}\le n-W(k)}(1+\sqrt{2})^{w_{0}}\sqrt{\frac{(n-W(k))!}{w_{0}!(n-W(k)-w_{0})!}}|1_{w_{0}}0_{n-W(k)-w_{0}}\rangle,
\end{equation}
\begin{equation}
|\tilde{\varphi}_{B}^{1}\rangle\equiv\sum_{0\le w_{1}\le W(k)}(-1)^{w_{1}}(1+\sqrt{2})^{-w_{1}}\sqrt{\frac{W(k)!}{w_{1}!(W(k)-w_{1})!}}|1_{w_{1}}0_{W(k)-w_{1}}\rangle,
\end{equation}
\begin{equation}
|\tilde{\varphi}_{A}^{2}\rangle\equiv\sum_{0\le w_{0}\le n-W(k)}(1-\sqrt{2})^{w_{0}}\sqrt{\frac{(n-W(k))!}{w_{0}!(n-W(k)-w_{0})!}}|1_{w_{0}}0_{n-W(k)-w_{0}}\rangle,
\end{equation}
\begin{equation}
|\tilde{\varphi}_{B}^{2}\rangle\equiv\sum_{0\le w_{1}\le W(k)}(-1)^{w_{1}}(1-\sqrt{2})^{-w_{1}}\sqrt{\frac{W(k)!}{w_{1}!(W(k)-w_{1})!}}|1_{w_{1}}0_{W(k)-w_{1}}\rangle,
\end{equation}
and normalize them using the following equations:
\begin{equation}
\sum_{0\le w_{0}\le n-W(k)}(1+\sqrt{2})^{2w_{0}}C_{n-W(k)}^{w_{0}}=(4+2\sqrt{2})^{n-W(k)},
\end{equation}
\begin{equation}
\sum_{0\le w_{1}\le W(k)}(1+\sqrt{2})^{-2w_{1}}C_{W(k)}^{w_{1}}=(4-2\sqrt{2})^{W(k)}.
\end{equation}
We remove the tilde notations for the normalized vectors, and the
eigenstate turns into
\begin{equation}
\begin{aligned}|\eta_{k}\rangle= & C\bigg[\frac{(1+\sqrt{2})^{W(k)-1}}{2\sqrt{2}}\sqrt{(4+2\sqrt{2})^{n-W(k)}}|\varphi_{A}^{1}\rangle\otimes\sqrt{(4-2\sqrt{2})^{W(k)}}|\varphi_{B}^{1}\rangle\\
 & -\frac{(1-\sqrt{2})^{W(k)-1}}{2\sqrt{2}}\sqrt{(4-2\sqrt{2})^{n-W(k)}}|\varphi_{A}^{2}\rangle\otimes\sqrt{(4+2\sqrt{2})^{W(k)}}|\varphi_{B}^{2}\rangle\bigg].
\end{aligned}
\end{equation}

The density matrix of part $A$ is
\begin{equation}
\begin{aligned}\rho_{A}=\rho_{\uparrow}= & \mathrm{tr_{B}}(|\eta_{k}\rangle\langle\eta_{k}|)\\
= & |\varphi_{A}^{1}\rangle\langle\varphi_{A}^{1}|C^{2}\frac{(1+\sqrt{2})^{2(W(k)-1)}}{8}(4+2\sqrt{2})^{n-W(k)}(4-2\sqrt{2})^{W(k)}\\
+ & |\varphi_{A}^{2}\rangle\langle\varphi_{A}^{2}|C^{2}\frac{(1-\sqrt{2})^{2(W(k)-1)}}{8}(4-2\sqrt{2})^{n-W(k)}(4+2\sqrt{2})^{W(k)},
\end{aligned}
\end{equation}
where $\langle\tilde{\varphi}_{A}^{1}|\tilde{\varphi}_{A}^{2}\rangle=\sum_{0\le w_{0}\le n-W(k)}(-1)^{w_{0}}C_{n-W(k)}^{w_{0}}=0$
and $C^{2}=\{2^{-3+n}(\sqrt{2}-1)^{2}[(2+\sqrt{2})^{n}+4(2-\sqrt{2})^{n-4}]\}^{-1}$.

Finally, we obtain
\begin{equation}
\begin{aligned}\rho_{\uparrow}=\rho_{A}= & |\varphi_{A}^{1}\rangle\langle\varphi_{A}^{1}|[1+(3-2\sqrt{2})^{n}(17+12\sqrt{2})]^{-1}+|\varphi_{A}^{2}\rangle\langle\varphi_{A}^{2}|[1+(3+2\sqrt{2})^{n}/(17+12\sqrt{2})]^{-1}\\
= & |\varphi_{A}^{1}\rangle\langle\varphi_{A}^{1}|[1+(3-2\sqrt{2})^{n-2}]^{-1}+|\varphi_{A}^{2}\rangle\langle\varphi_{A}^{2}|[1+(3+2\sqrt{2})^{n-2}]^{-1}.
\end{aligned}
.
\end{equation}
So the von Neumann entropy is given by
\begin{equation}
S(\rho_{\uparrow})=-\sum_{i=1,2}\nu_{i}{\rm \log}\nu_{i},\label{eq:entropy}
\end{equation}
in which $\nu_{1}=[1+(3-2\sqrt{2})^{n-2}]^{-1}$, $\nu_{2}=[1+(3+2\sqrt{2})^{n-2}]^{-1}$.

Similarly, for the $(k+1)$th eigenstate ($k$ is even),
\begin{equation}
\begin{aligned}|\eta_{k+1}\rangle= & C\sum_{\substack{0\le w_{1}\le W(k)\\
0\le w_{0}\le n-W(k)
}
}(-1)^{w_{1}}F(W(k)-w_{1}+w_{0})\sqrt{\frac{(n-W(k))!}{w_{0}!(n-W(k)-w_{0})!}}\sqrt{\frac{W(k)!}{w_{1}!(W(k)-w_{1})!}}\\
 & \times|1_{w_{0}}0_{n-W(k)-w_{0}}\rangle\otimes|1_{w_{1}}0_{W(k)-w_{1}}\rangle,
\end{aligned}
\end{equation}
where $C^{2}=\{\frac{1}{8}((4-2\sqrt{2})^{n}+2^{n}(2+\sqrt{2})^{n})\}^{-1}$.
The density matrix of part $A$ is
\begin{equation}
\begin{aligned}\rho_{A}=\rho_{\uparrow}= & \mathrm{tr_{B}}(|\eta_{k+1}\rangle\langle\eta_{k+1}|)\\
= & |\varphi_{A}^{1}\rangle\langle\varphi_{A}^{1}|C^{2}\frac{(1+\sqrt{2})^{2W(k)}}{8}(4+2\sqrt{2})^{n-W(k)}(4-2\sqrt{2})^{W(k)}\\
 & +|\varphi_{A}^{2}\rangle\langle\varphi_{A}^{2}|C^{2}\frac{(1-\sqrt{2})^{2W(k)}}{8}(4-2\sqrt{2})^{n-W(k)}(4+2\sqrt{2})^{W(k)}\\
= & |\varphi_{A}^{1}\rangle\langle\varphi_{A}^{1}|[1+(3-2\sqrt{2})^{n}]^{-1}+|\varphi_{A}^{2}\rangle\langle\varphi_{A}^{2}|[1+(3+2\sqrt{2})^{n}]^{-1}.
\end{aligned}
\end{equation}
So in Eq. \eqref{eq:entropy}, $\nu_{1}=[1+(3-2\sqrt{2})^{n}]^{-1}$,
$\nu_{2}=[1+(3+2\sqrt{2})^{n}]^{-1}$.

\twocolumngrid

\bibliography{ref1}

\end{document}